\documentclass[twocolumn,aps,superscriptaddress]{revtex4-1}
\usepackage{palatino}
\usepackage[latin1]{inputenc}
\usepackage{epsf}
\usepackage{amsmath,amssymb}
\usepackage{latexsym}
\usepackage{calc}
\usepackage{color}
\usepackage{shadow}
\usepackage{epsfig}
\usepackage[usenames,dvipsnames]{xcolor}

\def\bea{\begin{eqnarray}}
\def\eea{\end{eqnarray}}
\def\ben{\begin{equation}}
\def\een{\end{equation}}
\def\benu{\begin{enumerate}}
\def\enu{\end{enumerate}}

\def\sss{\scriptscriptstyle\rm}

\def\braket#1#2{\langle #1 \vert #2 \rangle}

\def\hatT{{\hat T}}

\def\1var{(\bx_1...\bx\N)}

\def\sss{\scriptscriptstyle\rm}
\def\x{_{\sss X}}
\def\c{_{\sss C}}
\def\s{_{\sss S}}
\def\xc{_{\sss XC}}

\def\H{_{\sss H}}

\def\S{^{\sss S}}
\def\NS{^{\sss NS}}

\def\br{{\bf r}}

\def\bx{{\br t}}

\begin{document}
\title{Time-Dependent Density Functional Theory Beyond Kohn-Sham Slater Determinants}
\author{Johanna I. Fuks}
\affiliation{Department of Physics and Astronomy, Hunter College and the Graduate Center of the City University of New York, 695 Park Avenue, New York, New York 10065, USA}
\author{S{\o}ren E. B. Nielsen}
\affiliation{Max Planck Institute for the Structure and Dynamics of Matter and Center for Free-Electron Laser Science \& Department of Physics, Luruper Chaussee 149, 22761 Hamburg, Germany}
\affiliation{Institut f\"ur Theoretische Physik, Universit\"at Innsbruck, Technikerstra{\ss}e 21A, A-6020 Innsbruck, Austria}
\author{Michael Ruggenthaler}
\affiliation{Max Planck Institute for the Structure and Dynamics of Matter and Center for Free-Electron Laser Science \& Department of Physics, Luruper Chaussee 149, 22761 Hamburg, Germany}
\affiliation{Institut f\"ur Theoretische Physik, Universit\"at Innsbruck, Technikerstra{\ss}e 21A, A-6020 Innsbruck, Austria}
\author{Neepa T. Maitra}
\affiliation{Department of Physics and Astronomy, Hunter College and the Graduate Center of the City University of New York, 695 Park Avenue, New York, New York 10065, USA}

\date{\today}

\begin{abstract} 
When running time-dependent density functional theory (TDDFT)
calculations for real-time simulations of non-equilibrium dynamics,
the user has a choice of initial Kohn-Sham state, and typically a
Slater determinant is used. We explore the impact of this choice on
the exchange-correlation potential when the physical system begins in
a 50:50 superposition of the ground and first-excited state of the
system. We investigate the possibility of
judiciously choosing a Kohn-Sham initial state that minimizes errors
when adiabatic functionals are used. We find that if the Kohn-Sham state is chosen to
have a configuration matching the one that dominates the interacting
state, this can be achieved for a finite time duration for some but not all such choices.
When the Kohn-Sham system does not begin in a Slater
determinant, we further argue that the conventional splitting of the
exchange-correlation potential into exchange and correlation parts has
limited value, and instead propose a decomposition into a
``single-particle'' contribution that we denote $v\xc\S$, and a
remainder. The single-particle contribution can be readily
computed as an explicit orbital-functional, reduces to exchange in the
Slater determinant case, and offers an alternative to the adiabatic approximation as a starting point 
for TDDFT approximations.
\end{abstract}
\maketitle 
\section{Introduction and Motivation}
Time-Dependent Density Functional Theory
(TDDFT)~\cite{Carstenbook,TDDFTbook12,RG84} is increasingly being used
beyond linear response applications, to study fully non-equilibrium
dynamics of electronic systems in real-time. Although TDDFT has had
much success in the linear response regime in calculations of spectra,
it is perhaps a more enticing tool for real-time dynamics when the
system is driven far from its ground-state, because of the
computational challenges in describing such dynamics for more than two electrons with alternative wavefunction-based methods. TDDFT
presents an attractive possibility due to its system-size scaling,
arising from its description of the dynamics in terms of a
non-interacting system of electrons evolving in a single-particle
potential. In practise the many-body effects hidden in this potential,
termed the exchange-correlation (xc) potential, must be approximated.
The Runge-Gross and the van Leeuwen \cite{L99} theorems reassure us that there exists an exact xc
potential for which the time-dependent one-body density of the interacting
system is exactly predicted.  It is therefore of much interest what
the features of this exact xc potential look like in order to develop
approximations that build in these features.

The exact xc potential at a given time $t$ is known to depend on the density
not just at time $t$ but also on its history, and it additionally depends on the
initial wavefunction of the physical system $\Psi(0)$, as well as the
choice of initial Kohn-Sham wavefunction $\Phi(0)$:
$v\xc[n;\Psi(0),\Phi(0)]$.  Typically however, in real-time TDDFT
simulations adiabatic approximations are used which depend only on the
instantaneous density: $v\xc^A[n;\Psi(0),\Phi(0)] = v\xc^{\rm
  g.s.}[n(t)]$. That is, they insert the instantaneous density into a
ground-state approximation. This approximation has provided useful
results for the interpretation and prediction of electron-dynamics in
a range of situations, including photovoltaic design~\cite{RFSR13},
high-harmonic generation in molecules~\cite{FB11}, coherent phonon
generation~\cite{SSYI12}, ultrafast demagnetization of
solids~\cite{EKDSG16}. However, in some cases there are complete
failures, e.g. Refs.~\cite{RN11,RN12c,HTPI14}, and studies on model systems have revealed
that there are large non-adiabatic features in the exact xc potential
that are missing in the approximations. These include step and peak
features that evolve in time~\cite{RG12,HRCLG13,HRDG14,EFRM12,FERM13,LFSEM14,LEM13}. These
dynamical features appear generically when the system has been pushed
out of equilibrium, whether or not there is an external field
present. They have been analyzed in terms of the local
acceleration across the system~\cite{EFRM12} and in terms of a
decomposition into kinetic and interaction xc
components~\cite{LFSEM14}. The latter generalizes a concept from the
ground-state that was pioneered by Evert Jan Baerends and his
group~\cite{BBS89,GLB94,GLB96,GB96}, which had provided valuable
insight into analyzing the structure of the ground-state xc potential
in atoms and molecules.

In this paper, we will focus on the initial-state dependence of these
non-adiabatic features, and on the xc potential in general. The
initial interacting state $\Psi(0)$ is given by the problem at hand,
but one has a choice of initial Kohn-Sham state $\Phi(0)$ in which to
begin the non-interacting Kohn-Sham evolution. 
The only restriction is that $\Phi(0)$ must have the same initial density and first time-derivative of the density as that of $\Psi(0)$.  
Almost always, an
initial single Slater determinant (SSD) of orbitals is chosen, even
when the initial interacting state is far from an SSD. For example, in
simulating photo-induced dynamics, where the true state is a singlet
single excitation of the molecule, a typical practise is simply to
raise one electron from the Kohn-Sham HOMO and place it in a virtual
orbital; spin-adaptation is done after the dynamics is run.
The true initial state on the other hand 
requires at least two Slater determinants for a minimal 
description. The exact xc potential for model systems with
different choices of initial Kohn-Sham states can vary significantly, and so the question arises: Is there an optimal choice of Kohn-Sham initial state, for which the exact xc potential has minimal non-adiabatic features, such that propagating this state using an adiabatic functional yields the smallest errors?  

Preliminary investigations considered initial
stationary excited states, and
suggested the ``best'' Kohn-Sham initial state when using an adiabatic
approximation has a configuration similar to the initial interacting
state~\cite{EM12}. That work was limited to studying the exact xc
potentials compared with the adiabatic ones  at the
initial time, while time-propagation was considered only using common
adiabatic approximations, since to find the exact xc potential for a Kohn-Sham state that has more than one occupied orbital is very demanding.
Refs.~\cite{NRL13,NRL14}, on the other hand, used the recently introduced
global fixed-point iteration
method~\cite{RL11,RGPL12,PR11,NRL13,RPL15,NRL14} to find the exact xc
potential in time for different choices of Kohn-Sham initial state for a given
 density of two particles interacting via a
cosine interaction on a ring. 

In this paper we go further, returning to
the more physical soft-Coulomb interaction used in Ref.~\cite{EM12},
and investigate the exact xc potential for different choices of
Kohn-Sham initial states as a function of time using the iteration
procedure of Ref.~\cite{NRL13,NRL14}, for field-free dynamics of a
superposition state. We compare with the adiabatically-exact (AE)
approximation, which is defined as the exact ground-state xc potential
evaluated on the exact density~\cite{TGK08}. This is the best possible
adiabatic approximation, and is what the intensive efforts in building
increasingly sophisticated approximations for ground-state problems
strive towards. We find that although the AE approximation 
is a better approximation for some choices of Kohn-Sham initial state than others, at least 
for short times, 
it is not particularly good for any of them for the strongly non-equilibrium
dynamics considered here. We expect therefore the AE propagation to be inaccurate.
Instead we introduce a different starting
point for TDDFT approximations, based on the time-dependent xc hole of the Kohn-Sham wavefunction. 
This is motivated by the decomposition of Refs.~\cite{RB09b, LFSEM14}, where the Kohn-Sham wavefunction is used to evaluate the interaction term, 
and yields an explicit orbital functional. We denote this approximation $v\xc\S$, and it contains some non-adiabatic contributions. 
 The exact xc potential is decomposed as $v\xc = v\xc\S + v\xc\NS$, which
we propose as the basis for a new approximation strategy which is particularly promising in the case of initial Kohn-Sham states that 
resemble the initial interacting state, going beyond the usual Kohn-Sham SSD.

\section{Model System and Different Initial States}
\label{sec:model}
The soft-Coulomb interaction~\cite{JES88} has been used extensively in strong-field
physics as well as in density-functional theory as it captures much of
the physics of real atoms and molecules. The particular Hamiltonian we
choose represents a model of the Helium atom prepared in a
superposition state.  We place two soft-Coulomb
interacting electrons in a one-dimensional (1D) soft-Coulomb well (atomic units are used throughout),
\ben
\hat{H} = \sum_{i=1,2}\left(-\frac{1}{2}\frac{d^2}{dx_i^2} - \frac{2}{\sqrt{x_i^2+1}}\right)+\frac{1}{\sqrt{|x_1-x_2|^2+1}},
\label{eq:H}
\een
prepared in a 50:50 superposition state of the ground and first-excited singlet state
of the interacting system, that is then allowed to evolve freely:
\ben
\Psi(0) = \frac{1}{\sqrt{2}}\left(\Psi_0 + \Psi_1\right),
\label{eq:Psi0}
\een
where $\Psi_i=\Psi_i(x_1,x_2)$ denote the many-body eigenstates of Hamiltonian~(\ref{eq:H}).
 The dynamics of the density,
\ben
n(x,t) = 2\int dx' \vert \Psi(x,x',t)\vert^2\;,
\label{eq:n}
\een  
therefore is a periodic sloshing inside the well, with period $2\pi/(E_1-E_0) = 11.788$ a.u.
We place the system in a  box with infinite walls at
$x= \pm L/2 = 7.5$au, giving zero boundary conditions at $\pm L/2$.

\subsection{Different choices of initial Kohn-Sham state}
We will study the features of the exact xc potential evaluated at this many-body density when different initial Kohn-Sham states are propagated. First, we define
\ben
\label{eq:Phi_a}
\Phi^{(a)}(0) \equiv \frac{1}{\sqrt{1+a^2}}\left(\Phi_0 + a\Phi_1\right)
\een
where 
\ben
\Phi_0(x_1,x_2) = \Phi^{\rm g.s.} = \phi_0^{(a)}(x_1)\phi_0^{(a)}(x_2)
\een
is the non-interacting ground-state of an ($(a)$-dependent) potential
and 
\ben
\Phi_1(x_1,x_2) =\frac{1}{\sqrt{2}}\left(\phi_0^{(a)}(x_1)\phi_1^{(a)}(x_2) +\phi_1^{(a)}(x_1)\phi_0^{(a)}(x_2) \right)
\een
is the first non-interacting singlet single-excitation of that same
potential; $\phi_0^{(a)}$ and $\phi_1^{(a)}$ are the ground and
first-excited orbitals of this potential. 
Clearly the orbitals and potential depend on the parameter
$a$, but such that in all cases $\Phi^{(a)}$ 
reproduces the same initial density and zero current-density as
$\Psi(0)$. We will consider different values for $a$, focussing particularly
on:

(i) $\Phi(0) = \Phi^{a=0}$. By requiring the initial density and zero
current-density match, we immediately obtain $\phi_0^{a=0} =
\sqrt{n(x,0)/2}$. This is the conventional choice for TDDFT, an SSD.
The exact xc potential in this case,
$v\xc[n;\Psi(0),\Phi^{a=0}]$ was studied before for this problem
 in Refs.~\cite{EFRM12,LFSEM14}, and
large non-adiabatic step and peak features were observed. They are
missing in all adiabatic approximations, and even the exact
ground-state xc potential i.e. the AE, looks completely different than the exact.  A point of
the present paper is to explore whether a different choice of initial
Kohn-Sham state can reduce these non-adiabatic features.

(ii) $\Phi(0) =\Phi^{a=1}$, which is a non-interacting analog of the
form of the true interacting state. Unlike in case (i), there
are many possible pairs of orbitals with this form that reproduce a
given density and current-density. The reason is that here we do not consider a ground state and hence do not have the Hohenberg-Kohn uniqueness theorem. 
Our choice here has a
relatively large overlap with the interacting state. See also (iii).

(iii) $\Phi(0) =\widetilde{\Phi}^{a=1}$, has the same form of (ii) but
with a different pair of orbitals that are eigenstates of quite a different
potential, and with a significantly smaller overlap with the true
interacting state.

(iv) $\Phi(0) = \Psi(0)$; the initial Kohn-Sham state is chosen
identical to the initial interacting state.  

In addition to the four states above, we will also show some results
for other $a$ but will not study these situations in so much detail.

\begin{figure}[h]
\begin{center}
\includegraphics[width=0.5\textwidth]{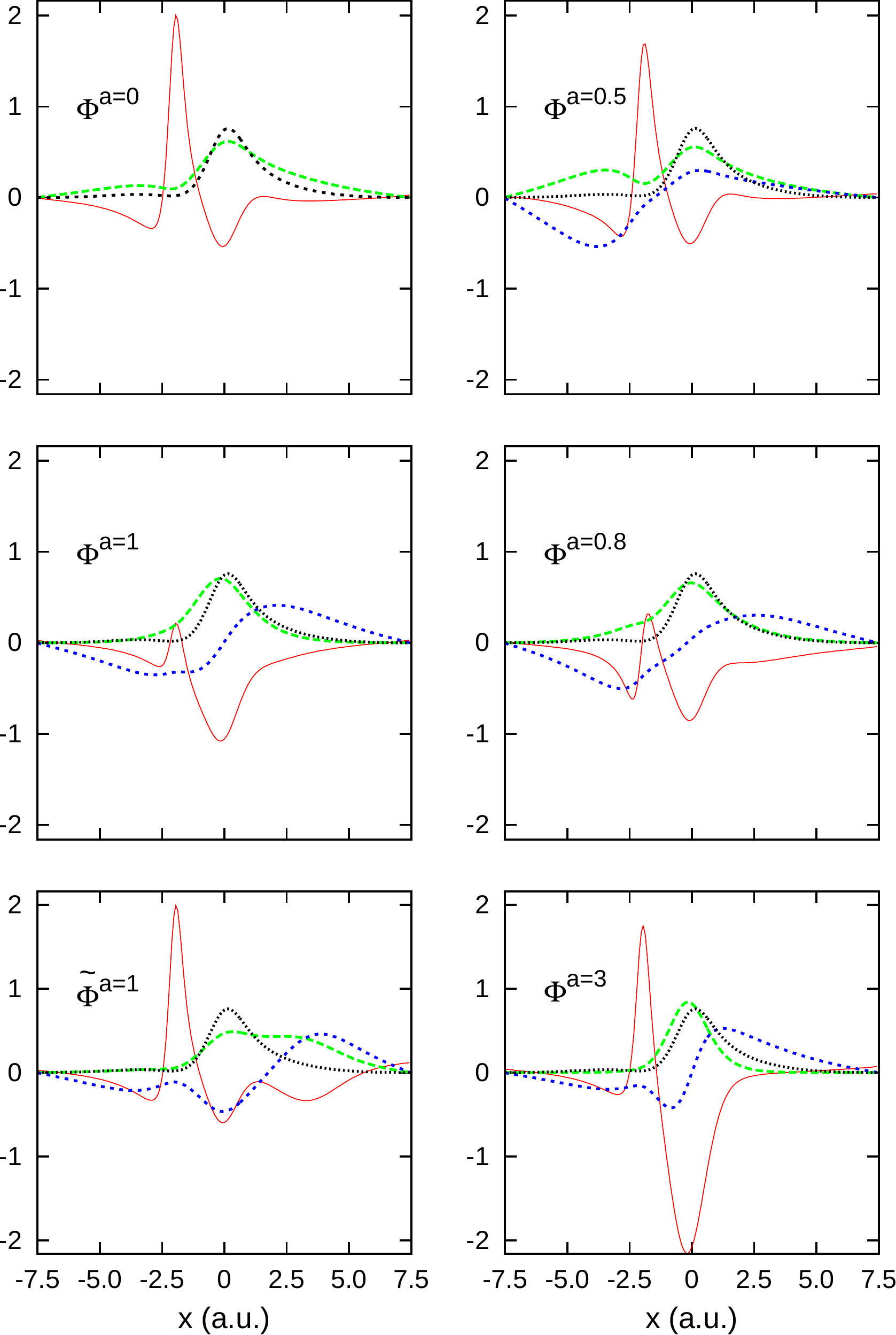}
\end{center}
\caption{Non-interacting potential $v$ (red, solid) for which the $\phi_0^{(a)}$ (green, long-dashed) and $\phi_1^{(a)}$ (blue, short-dashed) that form the initial KS wavefunction 
$\Phi^{(a)}(0)$, Eq.~(\ref{eq:Phi_a}), are eigenstates. In black, dotted, the initial density $n(x,0)$ is plotted.
}
\label{fig:initial-states}
\end{figure}

In Figure~\ref{fig:initial-states}, we plot the orbitals of these states, and the potentials in which they are eigenstates. 
While in the case $a=0$ we know the form of the initial Kohn-Sham orbitals and the corresponding Kohn-Sham potential analytically as a density-functional,
\bea
\label{eq:adexact}
v\s^{\rm g.s.}[n](x) = \frac{1}{2} \frac{\partial_x^2 \sqrt{n(x)}}{\sqrt{n(x)}},
\eea
 for $a\neq0$ we use an iterative method along the lines of Ref.~\cite{VLB1994}. An exposition of the method is given in appendix A.


Any of these states may be used to initiate a time-dependent Kohn-Sham calculation. 
If the exact xc potential was used, then propagating using $h\S$, 
\begin{align}
h\S  = \sum_{i=1,2}\Big(& -\frac{1}{2}\frac{d^2}{dx_i^2} -\frac{2}{\sqrt{x_i^2+1}} +v\H[n](x_i,t)\Big) 
\\
&+ v\xc[n;\Psi(0),\Phi(0)](x_i,t)\Big) \nonumber
\end{align}
where $v\H[n](x_i,t) = \int dx' n(x',t)/\sqrt{(x_i-x')^2+1}$ and $v\xc$ is different for each choice of initial Kohn-Sham state,  
then the exact time evolving density of the interacting system Eq.~(\ref{eq:n}) is reproduced.
In practise, the exact xc potential is unknown. How
closely the approximate functionals, in particular, adiabatic ones
that depend only on the instantaneous density, approximate the exact
xc potential gives an indication of how accurately the approximate functional
reproduces the density. Since adiabatic functionals neglect all
initial-state dependence (and dependence on the density's history),
then when evaluated on a given density, it gives the same potential for any of the choices of the Kohn-Sham initial state, unlike the exact xc potential.

In the next section we explain how we find the exact xc potential for a given choice of initial Kohn-Sham state.

\section{Finding the Exact xc Potential via the Global Fixed-Point Method}
\label{sec:global_fixed_point}
To find the exact xc potential for a given physical density evolution
is generally a non-trivial task. One must effectively find a one-body
potential in which non-interacting electrons evolve yielding the exact
physical density of the interacting system at each instant in time; even
whether, or when, such a potential exists is not a completely
closed question~\cite{RPL15}.
The non-adiabatic step and peak features that have been highlighted in Refs.~\cite{EFRM12,FERM13,LFSEM14} involve a particularly simple case: two electrons in 1D, where the Kohn-Sham initial state is taken to be a doubly-occupied orbital (as in (i) in the previous section), that evolves as $\phi(x,t) = \vert\phi(x,t)\vert e^{i\alpha(x,t)}$. 
The amplitude of this orbital is fixed from requiring the
instantaneous density is reproduced, $n(x,t) =
2\vert\phi(x,t)\vert^2$, while its phase follows from use of the
continuity equation in 1D: $\partial_t n(x,t) = -\partial_x j$ where $j$
is the one-body current, $j(x,t) = n(x,t) \partial_x\alpha(x,t)$. Thus the exact
Kohn-Sham orbital can be readily reconstructed from knowledge of
$n(x,t)$, and the Kohn-Sham equation inverted to find the Kohn-Sham
potential; subtracting the external and Hartree potentials yields the exact xc
potential. Despite the simple procedure (see also \cite{RNL13} for the periodic case), the resulting xc potential displays a wealth of interesting non-adiabatic features, absent in  ground-state potentials. 
For more electrons and more dimensions, a numerical iteration scheme is required instead, and has been performed for quasiparticle propagation  in a model one-dimensional 20-electron nano-wire~\cite{RG12} 
and for two electrons along a two-dimensional ring\cite{NRL14}. Sometimes lattice models can soothe the computational cost~\cite{SDS13,FM14,KPV11,VKFAB11}.

\subsection{Global Fixed Point Iteration Method}
Even for two electrons in 1D, the inversion to explore different possibilities
for Kohn-Sham initial states becomes non-trivial. As
soon as more than one Kohn-Sham orbital is involved, one must
guarantee not only that the time-dependent density is reproduced at
each instant in time from the sum-square of these orbitals, but also
that each orbital is evolving consistently in the same one-body potential.
In this work, we utilize the global fixed point iteration method
recently introduced in Refs.~\cite{RL11,RGPL12,PR11,NRL13,NRL14,RPL15} to find the
exact Kohn-Sham potential for the different choices of initial Kohn-Sham states. This method uses the fundamental equation of TDDFT, which is employed in the founding papers \cite{RG84,L99} and describes the divergence of the local forces (here in 1D for simplicity)
\ben
\label{eq:fund}
\partial_t^2 n(x,t) = \partial_x \left(n(x,t) \partial_x v\s(x,t)\right) -\partial_x Q\s(x,t),
\een
where $Q\s(x,t)=\braket{\Phi(t)}{\hat{Q}\s(x) \Phi(t)}$ is the internal force density of the Kohn-Sham system due to the 
momentum-stress forces $\hat{Q}\s(x) = -i [\hat{j}(x)  , \hatT]$ \cite{T2005}. In the case of the physical, interacting, system the internal force density 
also contains a contribution from the interaction forces, i.e., $\hat{Q}(x) = -i[\hat{j}(x), \hatT + \hat{W}]$ \cite{T2005}. 
If we prescribe the density $n(x,t)$ and fix an initial Kohn-Sham state $\Phi(0)$, then Eq.~(\ref{eq:fund}) becomes a non-linear equation for the 
external potential $v\s[n;\Phi(0)](x,t)$ that reproduces $n(x,t)$. The solution to this equation can then be found by an iteration procedure of the form
\bea
\label{eq:it}
-\partial_x \left(n(x,t) \partial_x v\s^{k+1}(x,t)\right) & =& \partial^2_t (n^k(x,t) - n(x,t)) \\
& -& \partial_x \left(n^k(x,t) \partial_x v\s^k(x,t)\right) , \nonumber 
\eea
where we used Eq.~(\ref{eq:fund}) to express $\partial_x Q\s^k(x,t)$ in terms of $v\s^k(x,t)$ and $n^k(x,t)$, the density produced by propagating $\Phi(0)$ with $v\s^k(x,t)$. In the case of zero boundary conditions on the wave functions we then solve for $v\s^{k+1}(x,t)$ by imposing that the potential needs to be even about the boundaries $x=\pm L/2$ \cite{NRL13,NRL14,RPL15}. We then iterate until convergence. [The actual numerical implementation uses a slightly different form of Eq.~(\ref{eq:it}) that also contains the current density and iterates for each individual time step. In this way numerical errors can be suppressed. For a detailed discussion of the algorithm the reader is refered to \cite{NRL14}]. This then allows us to determine the xc potential from
\bea
v\xc[n;\Psi(0),\Phi(0)] = v\s[n;\Phi(0)] - v - v\H[n],
\eea
where $v  = -2/\sqrt{x^2 +1}$ is explicitly given by the problem at hand. (As a functional $v = v[n;\Psi(0)]$, the potential that generates the prescribed density from $\Psi(0)$ in the interacting system, but we do not directly utilize this functional dependence.)

\section{Results: The exact time-dependent xc potential}
\label{sec:results_xcpotl}
Ref.~\cite{EFRM12} showed that for the Kohn-Sham non-interacting
system to exactly reproduce the density dynamics of the 50:50
superposition state when an SSD is used, the exact xc potential
develops large step and peak like features that are completely missed
by any adiabatic approximation, even the ``best'' adiabatic
approximation, the AE.   Can
these features be reduced if the Kohn-Sham system is allowed to go
beyond an SSD?

\begin{figure}[h]
\begin{center}
\includegraphics[width=0.5\textwidth, height=0.8\textwidth]{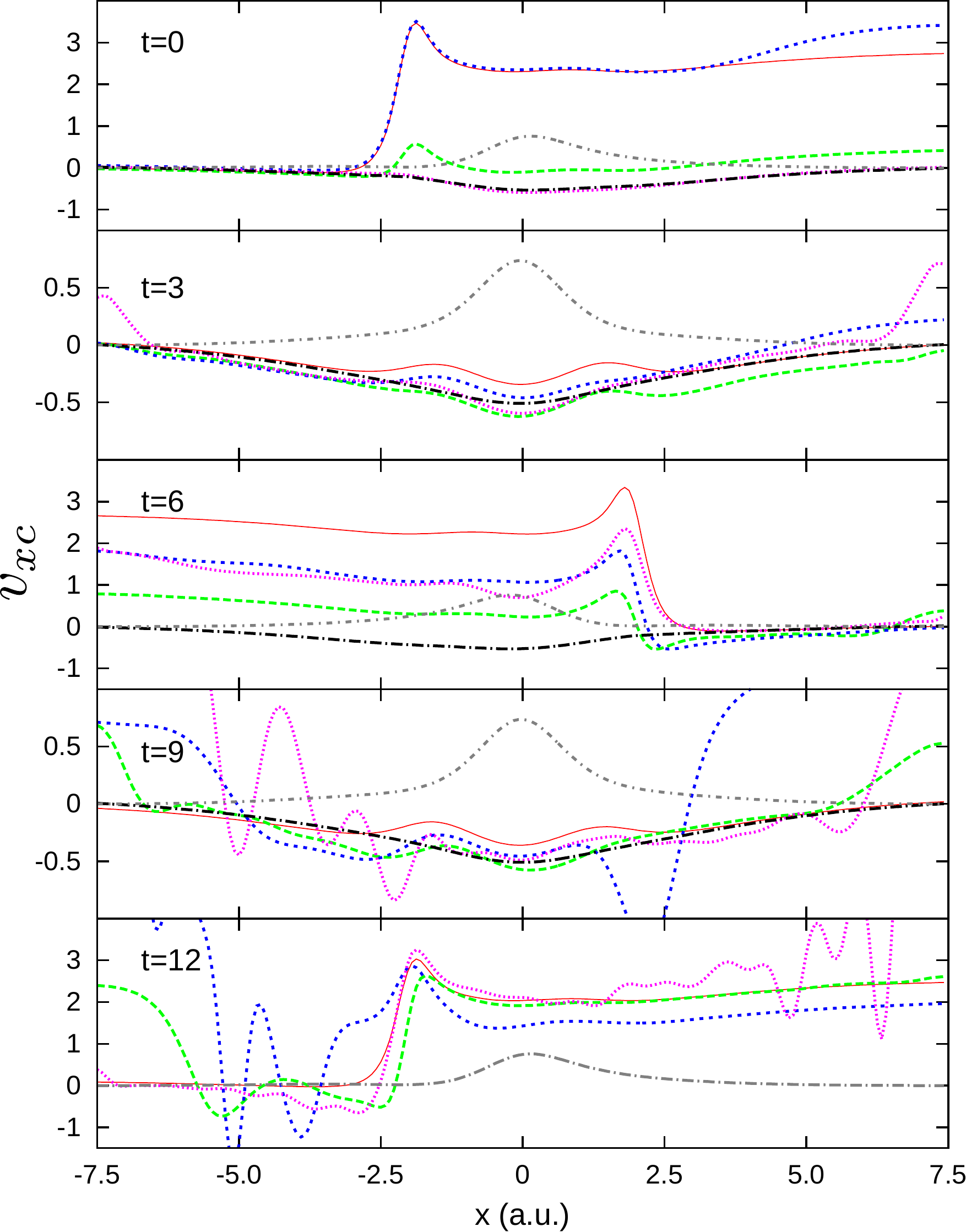}
\end{center}
\caption{$v\xc(x,t)$ for the different choices of KS initial state: $\Phi(0) ={\Phi}^{a=0}$ (red, solid), $\Phi(0) ={\Phi}^{a=1}$ (green, long-dashed),
$\Phi(0) =\widetilde{\Phi}^{a=1}$ (blue, short-dashed), $\Phi(0) =\Psi(0)$ (pink, dotted) and the adiabatically-exact $v\xc^{AE}$ (black, long-dashed-dot). 
The periodicity of the density $n(x,t)$, plotted as a guide in grey, short-dashed-dot, is $T=11.788$~a.u.
The oscillations seen in the potentials at longer times are a feature of our system, and not a numerical artifact;  they are converged with respect to 
spacing and time-step.
Please note the $y$-scale changes in each plot.
}
\label{fig:xcpot_diffIS1}
\end{figure}

\begin{figure}[h]
\begin{center}
\includegraphics[width=0.5\textwidth, height=0.8\textwidth]{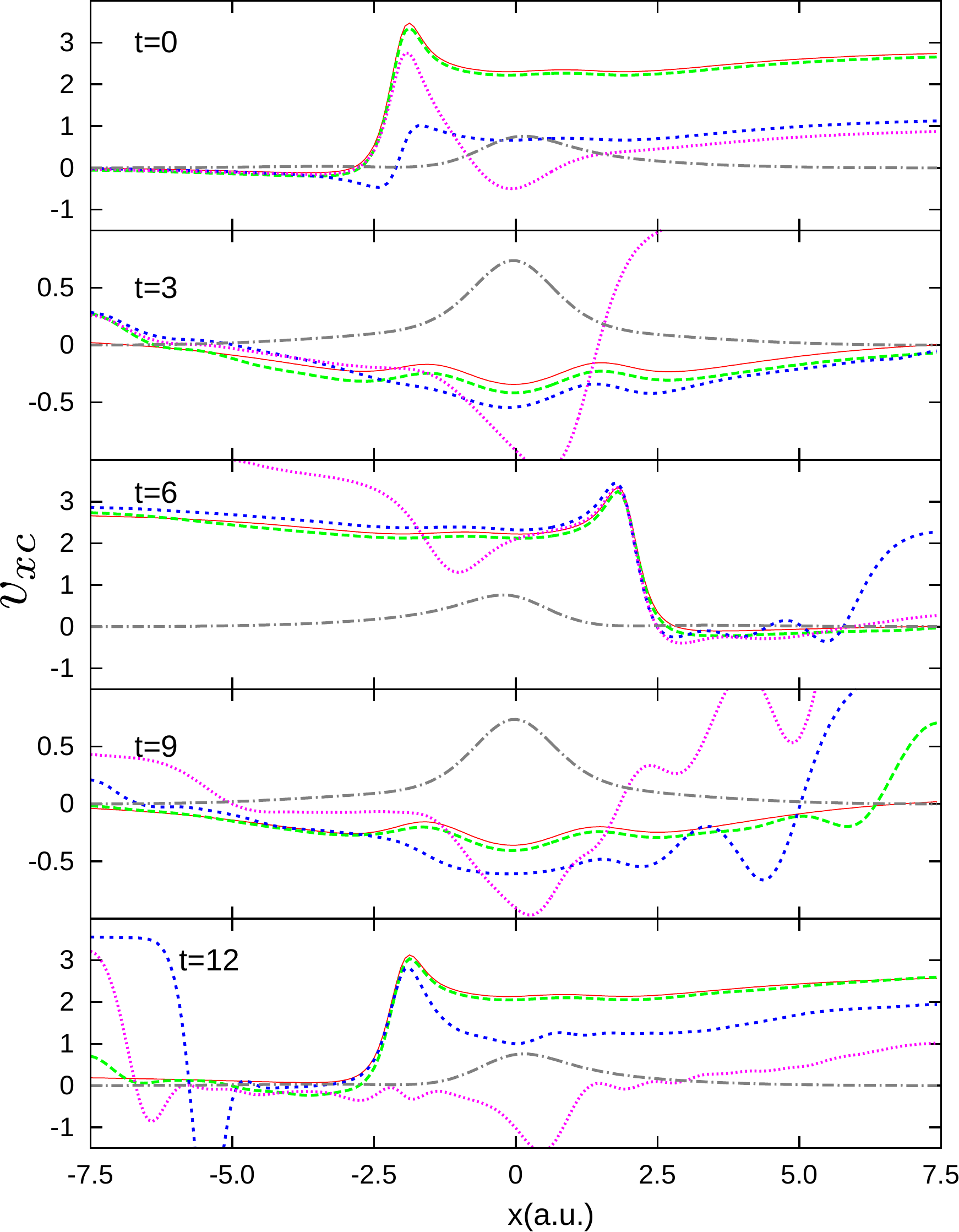}
\end{center}
\caption{$v\xc(x,t)$ for the different choices of KS initial state: $\Phi(0) ={\Phi}^{a=0}$ (red, solid), $\Phi(0) ={\Phi}^{a=0.5}$ (green, long-dashed),
$\Phi(0) ={\Phi}^{a=0.8}$ (blue, short-dashed), $\Phi(0) =\Phi^{a=3}$ (pink,dotted). The density $n(x,t)$ is plotted in grey, short-dashed-dot. } 
\label{fig:xcpot_diffIS2}
\end{figure}

\begin{figure}[h]
\begin{center}
\includegraphics[width=0.5\textwidth]{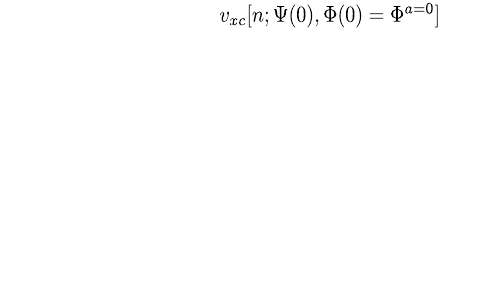}
\end{center}
\caption{$v\xc(x,t)$ for the SSD choice $\Phi(0) = \Phi^{a=0}$, shown up to time 15 a.u..} 
\label{fig:3Da0}
\end{figure}

\begin{figure}[h]
\begin{center}
\includegraphics[width=0.5\textwidth]{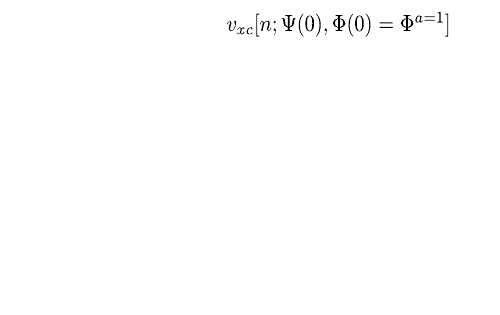}
\end{center}
\caption{$v\xc(x,t)$ for the choice $\Phi(0) = \Phi^{a=1}$, shown up to time 15 a.u..} 
\label{fig:3Da1}
\end{figure}

\begin{figure}[h]
\begin{center}
\includegraphics[width=0.5\textwidth]{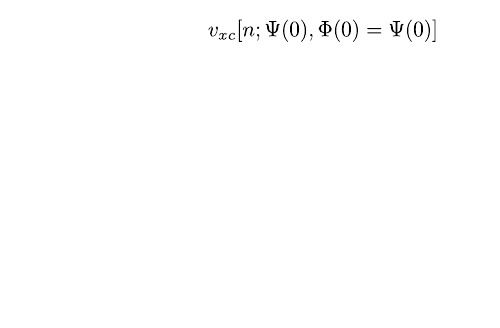}
\end{center}
\caption{$v\xc(x,t)$ for the choice $\Phi(0) = \Psi(0)$, shown up to time 15 a.u..} 
\label{fig:3Dpsi0}
\end{figure}

In Figures~\ref{fig:xcpot_diffIS1} and~\ref{fig:xcpot_diffIS2} we show
the exact xc potential $v\xc[n;\Psi(0),\Phi(0)](x,t)$ at various times
for the different choices of Kohn-Sham initial state detailed in
Section~\ref{sec:model}. 
The SSD case, $\Phi^{a=0}$ is shown in both plots as reference. 
The evolution of the xc potential for longer times 
is shown in Figures~\ref{fig:3Da0},~\ref{fig:3Da1}, and~\ref{fig:3Dpsi0}, for the KS initial state choices 
$a=0$, $a=1$ and $\Phi(0)=\Psi(0)$ respectively. 
  
We observe that the exact xc potential is hugely dependent on the
choice of initial state.  The prominent step feature at $t=0$ seen
when an SSD is used ($a=0$), which requires a non-adiabatic functional
to be captured, can be significantly reduced when other choices of
initial Kohn-Sham state are used.  
Note that although at $t=12$, in the middle region the step structure might 
appear more similar for several different choices of initial Kohn-Sham state 
than at earlier times, at later times they again become quite different, as clear in Figs~\ref{fig:3Da0},~\ref{fig:3Da1}, and~\ref{fig:3Dpsi0}.

{\underline{\it Observations for $\Phi^{a=1}$}}
Out of the few-determinant states
(i.e. all except for the $\Phi(0)=\Psi(0)$ case), the smallest xc effects are
observed at short times with state $\Phi^{a=1}$ which mimics most
closely the form of the interacting initial state with non-interacting
orbitals. The non-adiabatic features are smaller at short times,
although still present, as can be seen by comparing with the AE
plotted over the first period (since the density is periodic, and the
AE depends only on the instantaneous density, the AE also just repeats
over subsequent periods). For longer times though, the non-adiabatic
features can get large for $\Phi^{a=1}$, where at $t=12$ au, for example, we
see large oscillations in the tails and a step feature about the same
size as that of the SSD case.

{\underline{\it Observations for $\widetilde{\Phi}^{a=1}$}}
 However, simply having the right configuration is clearly not enough:
 at $t=0$ the potential for the state $\widetilde{\Phi}^{a=1}$ has
 just as prominent a step as the SSD (${\Phi}^{a=0}$) case and a very
 similar shape overall, especially on the left, with small gentle
 differences on the right except in the right-hand tail of the density
 where they differ more. Looking back to
 Fig.~\ref{fig:initial-states}, 
 the potentials in which
 the orbitals of $\widetilde{\Phi}^{a=1}$ and ${\Phi}^{a=0}$ are
 eigenstates are almost identical until the 
 initial density gets quite small on the right; in particular, they both possess a relatively sharp barrier
 near the minimum of the density on the left. The
$\widetilde{\Phi}^{a=1}$ orbitals are trapped between this large peak
 and the boundary at $L/2=7.5$ (see Fig.~\ref{fig:initial-states}).
 The occupied orbital of $v^{a=0}(x)$ (see upper left figure in Fig.~\ref{fig:initial-states}) is bound
 but with an energy lying close to the small barrier on its right,
 giving it a large tunneling tail. On the other hand, $v^{a=1}(x)$
 binds its lowest orbital more tightly to the position of its well,
 deeper than the cases of $v^{a=0}(x)$ and $\widetilde{v}^{a=1}$. Yet,
 these states all have the same one-particle density
 (Eq.~\ref{eq:n}).
Returning now to the xc potential that keeps the states evolving with the same density at the initial time, Fig.~\ref{fig:xcpot_diffIS1} top panel,
 we observe that the step feature on the left-hand-side is also very similar for the two states $\widetilde{\Phi}^{a=1}$  and $\Phi^{a=0}$. 
 This implies that the orbitals have a very similar local acceleration in that region~\cite{EFRM12}, which is not {\it a priori} obvious. 
 These two KS initial states have a very different nature, and their
overlaps with the
initial interacting state are also very different, $\langle\Psi(0)\vert\Phi^{a=0}(0)\rangle = 0.9202$ while $\langle\Psi(0)\vert\widetilde\Phi^{a=1}(0)\rangle = 0.7819$.
(In fact $\langle\Psi(0)\vert\Phi^{a=1}(0)\rangle = 0.9935$, much closer to $\langle\Psi(0)\vert\Phi^{a=0}(0)\rangle$, yet the resulting xc potential at
the initial time is  very different). 
At longer times, the potentials do
begin to differ significantly though as expected.  The large step in the potential for $\Phi^{a=0}$ dynamically oscillates in a periodic manner, while 
the potential for $\widetilde{\Phi}^{a=1}$ gets rather unruly quite quickly. 
Note that the periodicity of the density
carries over to the potential for the single-orbital case ($\Phi(0)=\Phi^{a=0}$) and the AE potential, due to the fact that
the single Kohn-Sham orbital depends simply and directly only on the density and
its first time-derivative (see Sec~\ref{sec:global_fixed_point}). This
is not the case for the other Kohn-Sham initial states, and their
periodicity is more complicated. Another striking example can be found in \cite{NRL13},
where the Kohn-Sham potential for a static density but with the interacting 
wave function as initial Kohn-Sham state is considered. The Kohn-Sham potential is strongly 
time-dependent since it needs to keep the static form of the density despite the rapidly
changing (multi-determinantal) Kohn-Sham wave function.

{\underline{\it Observations for the other few-determinant states}}
The potentials for the other few-determinant states in
Fig.~\ref{fig:xcpot_diffIS2} all have significant step features at the
initial time and develop large oscillatory features quite quickly. The
$\Phi^{a=0.5}$ case remains relatively close to the SSD case on the
scale shown, while the xc potential for $\Phi^{a=3}$ has large wells and
barriers throughout. The case of $\Phi^{a=0.8}$ is the one closest to
the ``best'' few-determinant state $\Phi^{a=1}$ shown on the left for
short times, but develops already a large step after half a period.

{\underline{\it Observations for $\Phi(0) = \Psi(0)$}}
The choice of the (many-determinant) state $\Phi(0)=\Psi(0)$ yields a
potential strikingly smaller than all the other potentials at $t=0$,
which is very close to the AE potential.  After already half a period
however, the potential, grows for this case too, along with steps and
peaks. Note that the oscillations seen after half a period are a feature of our system,
and not numerical artifacts; they are converged with respect to numerical parameters.

At this point we wish to stress that it is not that we necessarily are
striving to find the smallest xc potential; rather, we are trying to
find one with either the smallest non-adiabatic features such that the
usual adiabatic approximations would work best, {\it or} that is in
some other way somehow amenable to practical approximations. Of
course, the latter is not a very clearly-defined goal, however we
discuss a few approaches in the next section. For now, regarding the
AE approximation, we see that it is a very poor approximation for most
the choices of initial state we have studied, the exception being the case
$\Phi(0) = \Psi(0)$ at very short times (within a quarter of a
period). We will in fact find an even better approximation to the potential for $\Phi(0) = \Psi(0)$ for very short times in the next section.

\section{Decompositions and Approximations of the xc potential}
\label{sec:decomp}
Often when considering approximations, it is natural to decompose the
exact xc potential into parts that could be treated at different
levels of approximation.  We will discuss here four different types
of decompositions.

{\underline{\it Exchange and Correlation}}
The first is the common decomposition into exchange and correlation,
as has been traditionally done in ground-state density functional theory (DFT) and electronic
structure theory in general. Exchange implies interaction effects
beyond Hartree that account only for the Pauli exclusion principle; it
arises very naturally therefore in any theory that utilizes an SSD
reference, since an SSD takes care of the antisymmetry but does not
introduce any other interaction effects. For example, the ground-state
DFT (or Hartree-Fock) exchange energy is defined exactly as the
expectation value of the electron-electron operator in the Kohn-Sham
(or Hartree-Fock) SSD minus the Hartree energy. This decomposition has
affected the development of functional approximations. The
exchange energy is known exactly, is self-interaction-free, and dominates over correlation for atomic and molecular ground-states at equilibrium geometries. 
Because of this, some developers argue the exact exchange energy should be
used~\cite{G99}, while others~\cite{BG05} make the case that exchange and correlation should be approximated together to
make use of error cancellation, especially for the electron-pair bond in molecules where correlation becomes large.  Now, in the time-dependent theory a new
element arises: the possibility of using Kohn-Sham states that are not
SSDs throws into question the relevance of this separation into
exchange and correlation. If we were to define exchange via
interaction effects beyond Hartree introduced by considering the
interaction operator on a Kohn-Sham state that is not an SSD, then
clearly this includes more than just the effects from the Pauli
exclusion principle. If we preserve the meaning of exchange as
representing Pauli exclusion then, for two electrons in a spin-singlet as
we have, $v\x = -v\H/2$, independent of the choice of the initial Kohn-Sham state. The exchange potential therefore for any two-electron singlet dynamics is adiabatic and has the shape of a simple well ``under'' the density, with all the more interesting structure seen in Figs \ref{fig:xcpot_diffIS1} 
and \ref{fig:xcpot_diffIS2} appearing in the correlation potential. 

{\underline{\it Adiabatic and Non-Adiabatic}}
The second decomposition we discuss is the usual one made in TDDFT: into adiabatic (A)  and non-adiabatic (NA) (also termed dynamical) components, defined as
\ben
v\xc[n;\Psi(0),\Phi(0)]= v\xc^{\rm A}[n(t)]+ v\xc^{\rm NA}[n;\Psi(0),\Phi(0)]\;,
\label{eq:vxcdecompA_NA}
\een
where 
\ben
 v\xc^{\rm A}[n] = v\xc^{\rm g.s.}[n] = v\s^{\rm g.s.}[n] -v^{\rm g.s.}[n] - v\H[n].
\label{eq:vxcA}
\een
If the exact ground-state xc potential is used in Eq.~(\ref{eq:vxcA}),
this defines the AE potential, which we plotted in Fig.~ \ref{fig:xcpot_diffIS1} and discussed in previous section. In our specific case we have given $v\s^{\rm g.s.}[n]$ explicitly in Eq.~(\ref{eq:adexact}). For $v^{\rm g.s.}[n]$ we employed the iteration method outlined in appendix A. In practise, most
calculations in TDDFT use an adiabatic approximation, setting
$v\xc^{\rm NA} =0$ and choosing one of the myriad approximations
developed for the ground-state on the right-hand-side of
Eq.~(\ref{eq:vxcA}).  
Given that the ``best'' adiabatic approximation, the AE, is not close to the exact xc potential for any of the choices of initial Kohn-Sham
state except at very short times for the $\Phi(0) = \Psi(0)$ choice, it is unlikely that any adiabatic approximation will propagate accurately for this dynamics.

\begin{figure}[h]
\begin{center}
\includegraphics[width=0.5\textwidth,height=0.55\textwidth]{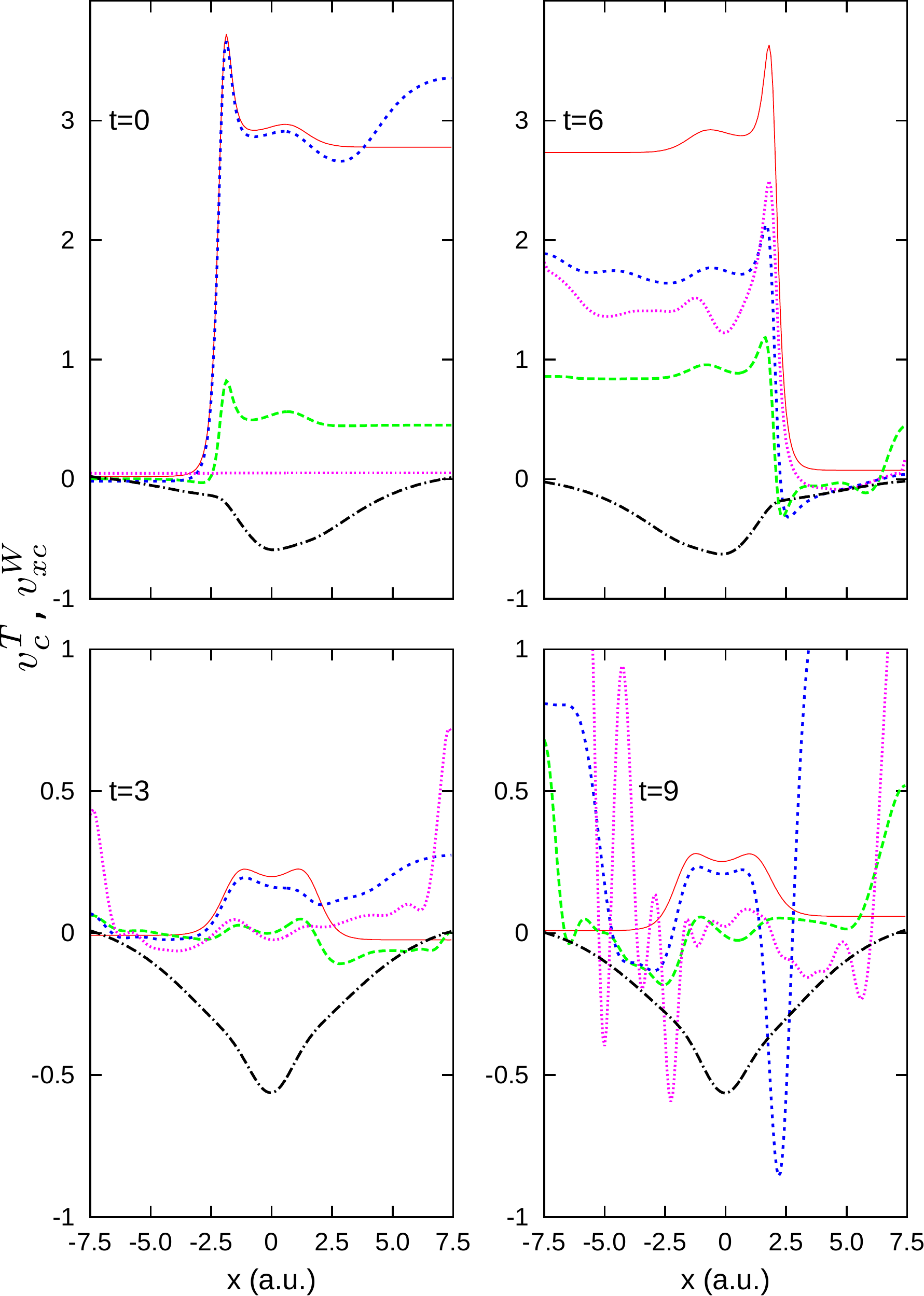}
\end{center}
\caption{
$v\c^T$ for the different choices of Kohn-Sham initial state: $\Phi(0) ={\Phi}^{a=0}$ (red, solid), $\Phi(0) ={\Phi}^{a=1}$ (green, long-dashed),
$\Phi(0) =\widetilde{\Phi}^{a=1}$ (blue, short-dashed), $\Phi(0) =\Psi(0)$ (pink, dotted). $v\xc^W$ is independent of the choice of initial Kohn-Sham state, shown in black (dashed-dot).
}
\label{fig:vcT_vxcW}
\end{figure}

{\underline{\it Kinetic and Interaction}}
A third decomposition is in terms of kinetic and interaction
components~\cite{L99,RB09b,LFSEM14}, which has only recently been
studied to understand better the nature of the xc potential, in
particular the non-adiabatic step and peak features that appear in
non-equilibrium dynamics~\cite{EFRM12,LFSEM14}. A similar
decomposition was pioneered by Evert Jan Baerends and his group for
the ground-state case~\cite{BBS89,GLB94,GLB96,GB96}, where it provided
an insightful analysis of step and peak features in ground-state
potentials of atoms and molecules. The ground-state xc potential is
expressed in terms of kinetic, hole, and response components.  The
kinetic and response components display peak and step features in
intershell regions in atoms, bonding regions in molecules, becoming
especially prominent at large bond-lengths, where they
  tend to be an indication of static correlation. In the time-dependent case, 
we define the interaction (W) and kinetic (T) terms 
\ben
v\xc = v\xc^W + v\c^T
\een
via Eq.~(\ref{eq:fund})
\begin{align}
 -\partial_x \left(n(x,t)\partial_x v\c^{T}(x,t)\right) = &  \partial_x \braket{\Psi(t)}{\hat{Q}\s(x) \Psi(t)} 
 \\
 & - \partial_x \braket{\Phi(t)}{\hat{Q}\s(x) \Phi(t)} \nonumber
\end{align}
\begin{align}
 -\partial_x \left(n(x,t)\partial_x v\xc^{W}(x,t)\right) & - \partial_x \left(n(x,t)\partial_x v\H(x,t) \right)
 \\
 & = - i \partial_x \braket{\Psi(t)}{[\hat{j}(x), \hat{W}] \Psi(t)} \nonumber
\end{align}
which can be solved for as
\bea
v\xc^W &=& \int^x dx'' \int n\xc(x',x'',t) \frac{\partial}{\partial x''} w(\vert x'-x'' \vert) dx' \\ \nonumber
\label{eq:vxcW}
\eea
\bea
v\c^T &=& \int^x \frac{1}{4n(x'',t)}\left(\frac{d}{dx'} - \frac{d}{dx''} \right)\left(\frac{d^2}{dx''^2}- \frac{d^2}{dx'^2}\right)\\
&&\left(\rho_1(x',x'',t) - \rho_{1,S}(x',x'',t) \right)\vert_{x'=x''} dx'' \nonumber
\label{eq:vcT}
\eea
where $n\xc$ is the xc hole, defined via the pair density, 
$P(x',x,t) = N(N-1)\sum_{\sigma_1..\sigma_N}\int \vert \Psi(x'\sigma_1,x\sigma_2,x_3\sigma_3..x_N\sigma_N; t) \vert^2 dx_3..dx_N 
 = n(x,t)\left(n(x',t) +n\xc(x',x,t)\right)$, and 
$
\rho_1(x',x,t) = N\sum_{\sigma_1..\sigma_N} \!\!\int dx_2...dx_N \Psi^*(x'\sigma_1,x_2\sigma_2...x_N\sigma_N;t)\times \\
\Psi(x\sigma_1,x_2\sigma_2 \dots x_N\sigma_N;t)$
is the spin-summed one-body density-matrix, and analogously for the Kohn-Sham system. 
The equations are written for the 1D case for simplicity. 
In Ref.~\cite{LFSEM14} it was shown that for the
SSD choice of Kohn-Sham initial state, $\Phi^{a=0}$, both $v\c^T$ and $v\xc^W$
display step structure although often that in $v\c^T$ dominates.
It was found in Ref.~\cite{LFSEM14} that the error made by the adiabatic approximation to $v\c^T$ is very large, 
typically larger than  that made to $v\xc^W$; the latter can be decomposed into the Coulomb potential of the xc hole and a correction, where the AE approximation does a fair job for the former. 
 Although $v\xc^W$ is independent of the choice of initial
Kohn-Sham state, $v\c^T$ is heavily dependent on
it. Figure~\ref{fig:vcT_vxcW}  
plots $v\xc^W$ and $v\c^T$ for 
Kohn-Sham initial state choices
$\Phi^{a=0},\Phi^{a=1},\widetilde{\Phi}^{a=1},\Psi(0)$ at several
times.
We see again that for the few-determinant
states, $v\c^T$ dominates the potential at the initial time and at half period ($t \sim 6$). 
For the choice $\Phi(0) = \Psi(0)$, $v\c^T$ is identically
zero at $t=0$; the entire xc potential is initially in the interaction term $v\xc^W$,
which is small. At later times however, $v\c^T$ becomes large also for this choice of KS initial state.

{\underline{\it Single-particle and Remainder}}
From the previous decomposition, a natural question then arises: if we approximate $v\xc^W$ by
inserting the instantaneous Kohn-Sham xc hole on the right-hand-side
of Eq~(\ref{eq:vxcW}), then how well does it do? If we made the analogous
approximation to $v\c^T$, approximating the true density-matrix by the Kohn-Sham one,
it would yield zero, so such an approach
would have to be used either if some other approximation for $v\c^T$
was used in conjunction, or if $v\c^T$ itself is very small. Indeed
the latter occurs when $\Phi(0) = \Psi(0)$ for very short times. 
At the initial time, it is clear from Eq.~(\ref{eq:vcT}) that when $\Psi(0) =\Phi(0)$, $v\xc^T = 0$ and $v\xc =v\xc^W$.
This observation motivates therefore a fourth decomposition:
\ben
v\xc = v\xc\S + v\xc\NS
\een
where the $v\xc\S$ represents the ``single-particle'' contribution
\ben
v\xc\S = \int^x dx''\int n\xc\S(x',x'',t) \frac{\partial}{\partial x''} w(\vert x'-x'' \vert) dx' 
\een
where $n\xc\S(x',x'',t)$ is defined via $P\S(x',x,t) \equiv N(N-1)\sum_{\sigma_1..\sigma_N}\int \vert \Phi(x'\sigma_1,x\sigma_2,x_3\sigma_3...x_N\sigma_N; t) \vert^2 dx_3..dx_N 
 = n(x,t)\left(n(x',t) +n\xc\S(x',x,t)\right)$
and $v\xc\NS$ represents the remainder, the non-single-particle
component.  In fact, $v\xc\S$, which may be viewed as an
orbital-dependent functional~\cite{RB09b}, is relatively simple to compute and to
propagate with, so if $v\xc\S$ is a good approximation to $v\xc$, then
this would represent a promising new direction for functional
development. Clearly, whether it is a good approximation or not
depends strongly on the choice of the Kohn-Sham initial state. If an
SSD is chosen, $v\xc\S$ reduces to the time-dependent exact-exchange
approximation. In the general case, it includes some correlation.  Our
previous plots suggest that, for the field-free evolution of the 50:50
superposition state with the chosen Kohn-Sham initial states we are
studying at present, although $v\xc\S$ may be a reasonable
approximation to $v\xc^W$ it will be a poor approximation to $v\xc$
because of the dominating $v\xc^T$ component, except for the
case $\Psi(0)=\Phi(0)$ where it will be very good at short times but likely not good at longer times. 

\begin{figure}[h]
\begin{center}
\includegraphics[width=0.5\textwidth,height=0.55\textwidth]{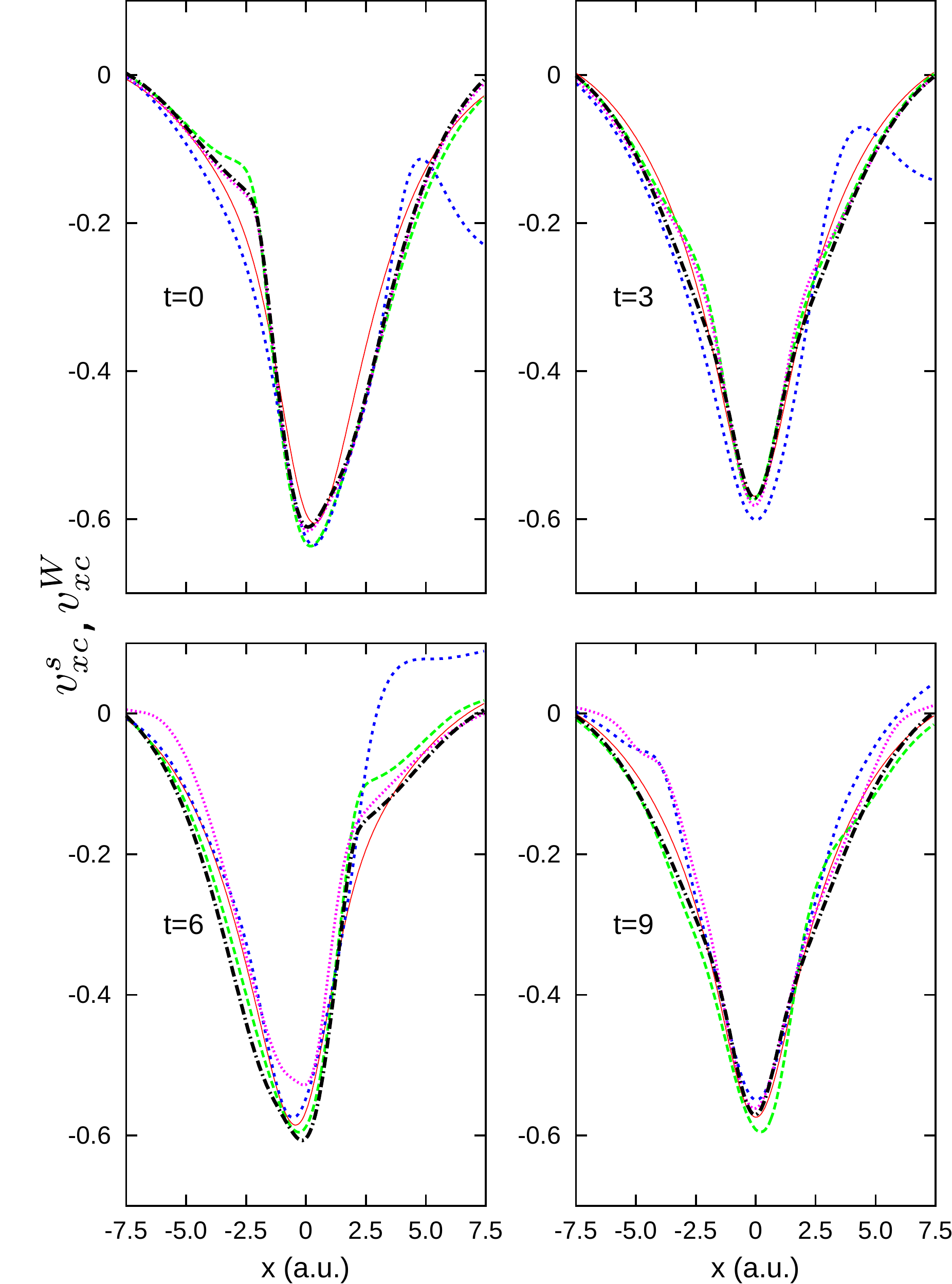}
\end{center}
\caption{$v\xc\S$ for the different choices of Kohn-Sham initial state: $\Phi(0) ={\Phi}^{a=0}$ (red, solid), $\Phi(0) ={\Phi}^{a=1}$ (green, long-dashed),
$\Phi(0) =\widetilde{\Phi}^{a=1}$ (blue, short-dashed), $\Phi(0) =\Psi(0)$ (pink , dotted) 
versus $v\xc^W$ (black, dashed-dot). Notice that $v\xc\S$ approximates well $v\xc^W$ for all choices of KS initial state, especially for the choice
$\Phi(0)=\Psi(0)$ at early times.
}
\label{fig:vxcS_vxcW_AE}
\end{figure}

\begin{figure}[h]
\begin{center}
\includegraphics[width=0.5\textwidth,height=0.55\textwidth]{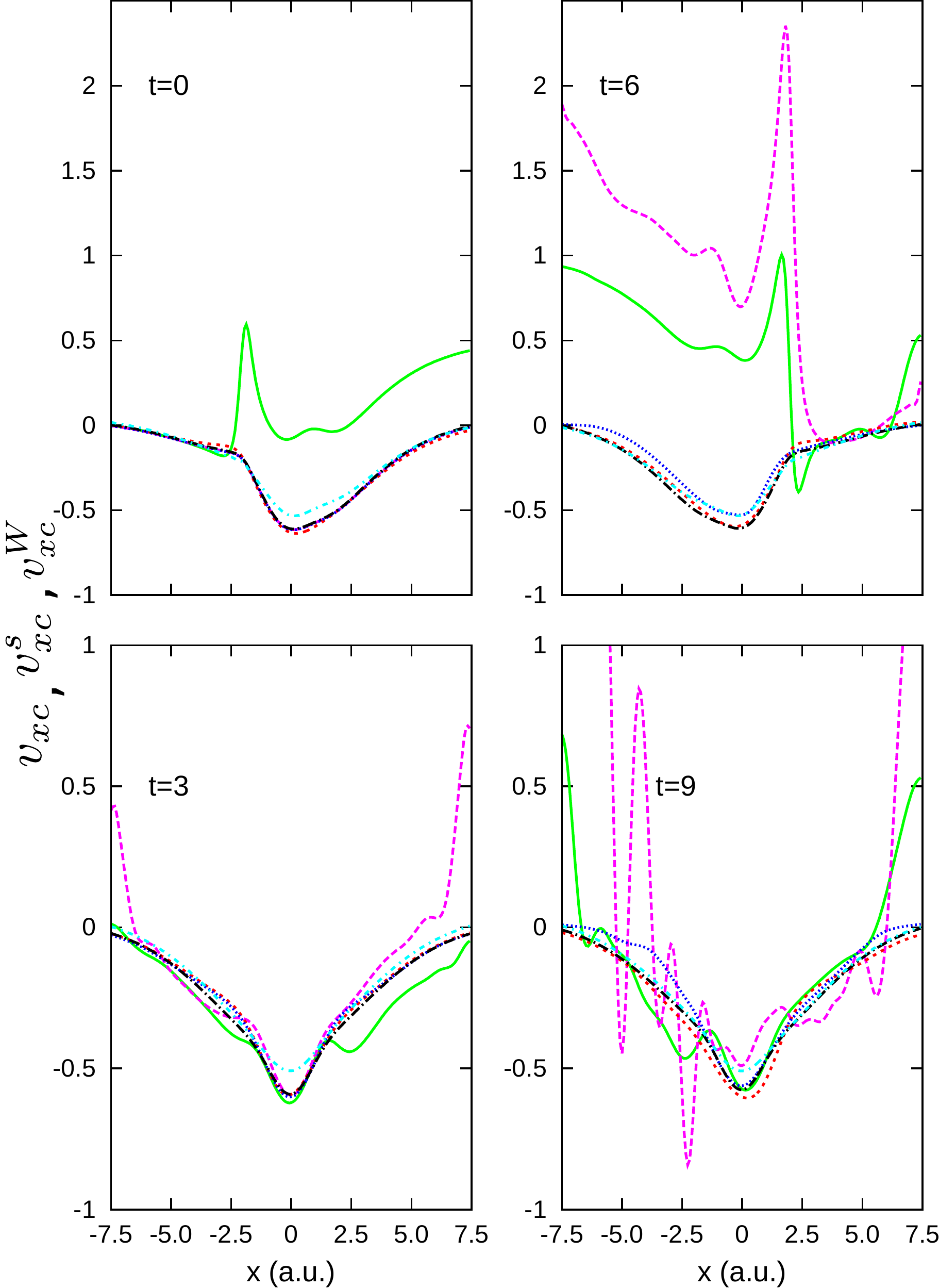}
\end{center}
\caption{$v\xc$ and $v\xc\S$ for the choices of KS initial state:  $\Phi(0) ={\Phi}^{a=1}$ 
(green solid; red short-dashed),$\Phi(0) =\Psi(0)$ (pink long-dashed; blue dotted) along with the AE xc potential (light-blue, dot-dashed) 
and $v\xc^W$ (black, dot-long-dashed).
}
\label{fig:vxc_vxcS_vxcW}
\end{figure}

Figures~\ref{fig:vxcS_vxcW_AE} and~\ref{fig:vxc_vxcS_vxcW} verify this. 
Figure~\ref{fig:vxcS_vxcW_AE} shows that $v\xc\S$ is an excellent
approximation to $v\xc^W$ for the case $\Phi(0) = \Psi(0)$, especially
at earlier times. It is also not a bad approximation for the other
initial state choices. 
Figure~\ref{fig:vxc_vxcS_vxcW} compares then the approximation $v\xc\S$ to the full $v\xc$ 
for the cases in which $v\c^T$ is the smallest, $\Phi(0) = \Psi(0)$ and $\Phi(0) = \Phi^{a=1}$, along with the AE xc potential,
and bears out the expectations of the previous paragraph.

The integrated nature of $v\xc^W$ (see Eq.~\ref{eq:vxcW}) makes it much more forgiving to approximate xc holes, than the $v\c^T$ is to the density-matrix; the high number of derivatives in Eq.~\ref{eq:vcT} make $v\c^T$ very sensitive. 


\section{Conclusions and Outlook}
Initial-state dependence is a subtle aspect of the exact xc potential
of TDDFT that has no precedent in ground-state DFT. Here we have
explored the possibility of exploiting it to ease the job of the xc
functional, i.e. to investigate whether, for a given initial state of
the interacting problem, there is a ``most suitable'' Kohn-Sham
initial state to use such that features of the xc functional are
simpler to approximate. We note that most calculations today assume an
SSD for the Kohn-Sham system, but the theory has no such restriction.
We have focussed only on one case, field-free evolution of a 50:50
superposition of a ground and first-excited state of a two-electron
problem. This is, arguably, one of the most difficult cases for
conventional approximations in TDDFT, given the extreme
non-equilibrium nature of the dynamics. At the same time, although the
results here are discussed specifically for this particular evolution,
they have consequences also for dynamics driven by external fields
when beginning in the ground state, since such superpositions can be
reached in the interacting system, yet the Kohn-Sham state at all
times is restricted to whatever configuration it began in.

Earlier work had shown that if an SSD is chosen as the Kohn-Sham
state, large dynamical features, such as steps and peaks, appear in
the potential from the very onset of the dynamics.  These are
completely missed by any adiabatic approximation and have a dependence
on the density that is non-local in space and time. It has remained so
far a challenge to derive approximate density-functionals with memory
that incorporate these features, and so adds much  motivation to the
complementary path we investigate here, whether one can exploit
initial-state dependence to reduce their size. 

We have found that if the Kohn-Sham state is chosen to also have the
form of a 50:50 superposition state, then one can find such a state
for which the non-adiabatic step and peak features are small at
short times. But at longer times (in our case, within one period
of the dynamics), the non-adiabatic features can get large, as large
as the SSD choice.  However simply having the right configuration is
not enough; other initial states with this configuration can be found
for which the non-adiabatic features are again large even at short
times. Having said that, perhaps there is another Kohn-Sham state of
this form that we just did not find, that has actually smaller
non-adiabatic features.  If we instead consider the correlated
many-body interacting state as the initial Kohn-Sham state, then we
find that at very short times, the adiabatic approximation becomes a 
very accurate approximation to the xc potential, however even after
half a period, dynamical step and peak features reappear that are
larger than the best few-determinant choice. We can conclude that for
very short times, the choice of $\Phi(0) = \Psi(0)$ is best if one
must use an adiabatic approximation, while at longer but still short
times (on the time-scale of the dynamics), a judicious choice of a
Kohn-Sham initial state with the same configuration as the interacting
state is likely a better choice. 
This particular conclusion is not universally true, e.g. if the initial interacting state is a ground state, 
then the adiabatic approximation works best at short times if the Kohn-Sham state is a non-interacting ground-state, and the AE becomes then exact at the initial time~\cite{EM12}.

We propose a new direction for developing approximations for the xc
potential in cases where the Kohn-Sham state is chosen generally
(i.e. not restricted to an SSD). This is based on decomposing the
exact expression for the xc potential arising from the force-balance
equation of TDDFT into a single-particle component and the
remainder. One can express the exact xc potential as two
contributions, one is a kinetic term, and the other is an interaction
term.  The single-particle component, $v\xc\S$ has only terms arising
from the interaction, but evaluated using the Kohn-Sham xc hole; it
reduces therefore to exchange in the SSD limit, but in the general
case, includes correlation components too. It is straightforward to
compute as an orbital functional, and we found that, for initial
Kohn-Sham states judiciously chosen as discussed above, it gives an
extremely good approximation to the exact interaction term, and a very
good approximation to the full xc potential for short times. At longer
times, the kinetic term becomes important and $v\xc\S$ becomes
inaccurate. Future work involves focussing then on developing new
approximations for the kinetic component of the xc potential while
utilizing $v\xc\S$ for the interaction component.

\begin{acknowledgments}
Financial support from the National Science Foundation
CHE-1152784 (N.T.M), and Department of Energy, Office of Basic Energy
Sciences, Division of Chemical Sciences, Geosciences and Biosciences
under Award DE-SC0008623 (for J.I.F) are gratefully acknowledged. S.E.B.N. and M.R. acknowledge financial support by the FWF (Austrian Science Fund) through the project P 25739-N27.
\end{acknowledgments}

\appendix

\section{Finding adiabatic potentials and eigenstates for a given density and current}

To find a potential $v(x)$ for a given form of the Hamiltonian, i.e., interacting or non-interacting, that has an eigenstate which reproduces a 
given density $n(x)$ and current density $j(x)=0$, we employ an iterative method along the lines of van Leeuwen \cite{VLB1994}.
The iteration steps read
\bea
\label{eq:gsiterate}
v_{k+1}(x) = v_{k}(x) + \delta\left(\frac{n_k(x)}{n(x)} - 1\right), 
\eea
where $n_k(x)$ is the density of the eigenstate of $v_{k}(x)$ and we choose $\delta>0$ small enough.
We see that the procedure in each iteration simply increases the potential where the density is too large,
to push away density, and decreases the potential where the density is too small, to collect more density.
If we prescribe 
a ground-state density (no nodes) and use the ground-state density $n_{k}(x)$ of $v_k(x)$ in each iteration then, if the procedure converges, 
the solution is a unique ground-state due to the Hohenberg-Kohn uniqueness theorem. In this way we can determine the ground-state potentials 
$v^{\rm g.s.}[n](x)$ (see Eq.~(\ref{eq:vxcA}) and discussion thereafter) in the interacting case. This would also work in the non-interacting case, but in our two-particle spin-singlet case the 
ground-state potential in the non-interacting case is known explicitly and hence we can directly use Eq.~(\ref{eq:adexact}).
\\
On the other hand, if we restrict the form of our Kohn-Sham wave function $\Phi^{(a \neq 0)}$ in the above iteration scheme and thus have
\ben
n^{(a)}(x) = \frac{(2+a^2)\phi_0(x)^2 + 2\sqrt{2}a\phi_0(x)\phi_1(x) + a^2 \phi_1(x)^2  }{(1+a^2)},
\label{eq:density_A}
\een
we are constructing an excited state of a non-interacting problem. Hence, if we converge, we can no longer rely on the Hohenberg-Kohn uniqueness theorem and consequently we can find several solutions, such as in the case $a=1$ (see cases (ii) and (iii) in section II).



\addcontentsline{toc}{section}{References}
\bibliography{./ref}

\begin{thebibliography}{41}%
\makeatletter
\providecommand \@ifxundefined [1]{%
 \@ifx{#1\undefined}
}%
\providecommand \@ifnum [1]{%
 \ifnum #1\expandafter \@firstoftwo
 \else \expandafter \@secondoftwo
 \fi
}%
\providecommand \@ifx [1]{%
 \ifx #1\expandafter \@firstoftwo
 \else \expandafter \@secondoftwo
 \fi
}%
\providecommand \natexlab [1]{#1}%
\providecommand \enquote  [1]{``#1''}%
\providecommand \bibnamefont  [1]{#1}%
\providecommand \bibfnamefont [1]{#1}%
\providecommand \citenamefont [1]{#1}%
\providecommand \href@noop [0]{\@secondoftwo}%
\providecommand \href [0]{\begingroup \@sanitize@url \@href}%
\providecommand \@href[1]{\@@startlink{#1}\@@href}%
\providecommand \@@href[1]{\endgroup#1\@@endlink}%
\providecommand \@sanitize@url [0]{\catcode `\\12\catcode `\$12\catcode
  `\&12\catcode `\#12\catcode `\^12\catcode `\_12\catcode `\%12\relax}%
\providecommand \@@startlink[1]{}%
\providecommand \@@endlink[0]{}%
\providecommand \url  [0]{\begingroup\@sanitize@url \@url }%
\providecommand \@url [1]{\endgroup\@href {#1}{\urlprefix }}%
\providecommand \urlprefix  [0]{URL }%
\providecommand \Eprint [0]{\href }%
\providecommand \doibase [0]{http://dx.doi.org/}%
\providecommand \selectlanguage [0]{\@gobble}%
\providecommand \bibinfo  [0]{\@secondoftwo}%
\providecommand \bibfield  [0]{\@secondoftwo}%
\providecommand \translation [1]{[#1]}%
\providecommand \BibitemOpen [0]{}%
\providecommand \bibitemStop [0]{}%
\providecommand \bibitemNoStop [0]{.\EOS\space}%
\providecommand \EOS [0]{\spacefactor3000\relax}%
\providecommand \BibitemShut  [1]{\csname bibitem#1\endcsname}%
\let\auto@bib@innerbib\@empty
\bibitem [{\citenamefont {Ullrich}(2011)}]{Carstenbook}%
  \BibitemOpen
  \bibfield  {author} {\bibinfo {author} {\bibfnamefont {C.~A.}\ \bibnamefont
  {Ullrich}},\ }\href@noop {} {\emph {\bibinfo {title} {Time-dependent
  density-functional theory: concepts and applications}}}\ (\bibinfo
  {publisher} {Oxford University Press},\ \bibinfo {year} {2011})\BibitemShut
  {NoStop}%
\bibitem [{\citenamefont {Marques}\ \emph {et~al.}(2012)\citenamefont
  {Marques}, \citenamefont {Maitra}, \citenamefont {Nogueira}, \citenamefont
  {Gross},\ and\ \citenamefont {Rubio}}]{TDDFTbook12}%
  \BibitemOpen
  \bibinfo {editor} {\bibfnamefont {M.~A.}\ \bibnamefont {Marques}}, \bibinfo
  {editor} {\bibfnamefont {N.~T.}\ \bibnamefont {Maitra}}, \bibinfo {editor}
  {\bibfnamefont {F.~M.}\ \bibnamefont {Nogueira}}, \bibinfo {editor}
  {\bibfnamefont {E.~K.}\ \bibnamefont {Gross}}, \ and\ \bibinfo {editor}
  {\bibfnamefont {A.}~\bibnamefont {Rubio}},\ eds.,\ \href@noop {} {\emph
  {\bibinfo {title} {Fundamentals of time-dependent density functional
  theory}}},\ Vol.\ \bibinfo {volume} {837}\ (\bibinfo  {publisher}
  {Springer},\ \bibinfo {year} {2012})\BibitemShut {NoStop}%
\bibitem [{\citenamefont {Runge}\ and\ \citenamefont {Gross}(1984)}]{RG84}%
  \BibitemOpen
  \bibfield  {author} {\bibinfo {author} {\bibfnamefont {E.}~\bibnamefont
  {Runge}}\ and\ \bibinfo {author} {\bibfnamefont {E.~K.~U.}\ \bibnamefont
  {Gross}},\ }\href {\doibase 10.1103/PhysRevLett.52.997} {\bibfield  {journal}
  {\bibinfo  {journal} {Phys. Rev. Lett.}\ }\textbf {\bibinfo {volume} {52}},\
  \bibinfo {pages} {997} (\bibinfo {year} {1984})}\BibitemShut {NoStop}%
\bibitem [{\citenamefont {van Leeuwen}(1999)}]{L99}%
  \BibitemOpen
  \bibfield  {author} {\bibinfo {author} {\bibfnamefont {R.}~\bibnamefont {van
  Leeuwen}},\ }\href {\doibase 10.1103/PhysRevLett.82.3863} {\bibfield
  {journal} {\bibinfo  {journal} {Phys. Rev. Lett.}\ }\textbf {\bibinfo
  {volume} {82}},\ \bibinfo {pages} {3863} (\bibinfo {year}
  {1999})}\BibitemShut {NoStop}%
\bibitem [{\citenamefont {Rozzi}\ \emph {et~al.}(2013)\citenamefont {Rozzi},
  \citenamefont {Falke}, \citenamefont {Spallanzani}, \citenamefont {Rubio},
  \citenamefont {Molinari}, \citenamefont {Brida}, \citenamefont {Maiuri},
  \citenamefont {Cerullo}, \citenamefont {Schramm}, \citenamefont
  {Christoffers} \emph {et~al.}}]{RFSR13}%
  \BibitemOpen
  \bibfield  {author} {\bibinfo {author} {\bibfnamefont {C.~A.}\ \bibnamefont
  {Rozzi}}, \bibinfo {author} {\bibfnamefont {S.~M.}\ \bibnamefont {Falke}},
  \bibinfo {author} {\bibfnamefont {N.}~\bibnamefont {Spallanzani}}, \bibinfo
  {author} {\bibfnamefont {A.}~\bibnamefont {Rubio}}, \bibinfo {author}
  {\bibfnamefont {E.}~\bibnamefont {Molinari}}, \bibinfo {author}
  {\bibfnamefont {D.}~\bibnamefont {Brida}}, \bibinfo {author} {\bibfnamefont
  {M.}~\bibnamefont {Maiuri}}, \bibinfo {author} {\bibfnamefont
  {G.}~\bibnamefont {Cerullo}}, \bibinfo {author} {\bibfnamefont
  {H.}~\bibnamefont {Schramm}}, \bibinfo {author} {\bibfnamefont
  {J.}~\bibnamefont {Christoffers}},  \emph {et~al.},\ }\href@noop {}
  {\bibfield  {journal} {\bibinfo  {journal} {Nature Communications}\ }\textbf
  {\bibinfo {volume} {4}},\ \bibinfo {pages} {1602} (\bibinfo {year}
  {2013})}\BibitemShut {NoStop}%
\bibitem [{\citenamefont {Penka~Fowe}\ and\ \citenamefont
  {Bandrauk}(2011)}]{FB11}%
  \BibitemOpen
  \bibfield  {author} {\bibinfo {author} {\bibfnamefont {E.}~\bibnamefont
  {Penka~Fowe}}\ and\ \bibinfo {author} {\bibfnamefont {A.~D.}\ \bibnamefont
  {Bandrauk}},\ }\href {\doibase 10.1103/PhysRevA.84.035402} {\bibfield
  {journal} {\bibinfo  {journal} {Phys. Rev. A}\ }\textbf {\bibinfo {volume}
  {84}},\ \bibinfo {pages} {035402} (\bibinfo {year} {2011})}\BibitemShut
  {NoStop}%
\bibitem [{\citenamefont {Shinohara}\ \emph {et~al.}(2012)\citenamefont
  {Shinohara}, \citenamefont {Sato}, \citenamefont {Yabana}, \citenamefont
  {Iwata}, \citenamefont {T.Otobe},\ and\ \citenamefont {Bertsch}}]{SSYI12}%
  \BibitemOpen
  \bibfield  {author} {\bibinfo {author} {\bibfnamefont {Y.}~\bibnamefont
  {Shinohara}}, \bibinfo {author} {\bibfnamefont {S.}~\bibnamefont {Sato}},
  \bibinfo {author} {\bibfnamefont {K.}~\bibnamefont {Yabana}}, \bibinfo
  {author} {\bibfnamefont {J.-I.}\ \bibnamefont {Iwata}}, \bibinfo {author}
  {\bibnamefont {T.Otobe}}, \ and\ \bibinfo {author} {\bibfnamefont {G.~F.}\
  \bibnamefont {Bertsch}},\ }\href@noop {} {\bibfield  {journal} {\bibinfo
  {journal} {J. Chem. Phys}\ }\textbf {\bibinfo {volume} {137}},\ \bibinfo
  {pages} {22A527} (\bibinfo {year} {2012})}\BibitemShut {NoStop}%
\bibitem [{\citenamefont {Elliott}\ \emph {et~al.}(2016)\citenamefont
  {Elliott}, \citenamefont {Krieger}, \citenamefont {Dewhurst}, \citenamefont
  {Sharma},\ and\ \citenamefont {Gross}}]{EKDSG16}%
  \BibitemOpen
  \bibfield  {author} {\bibinfo {author} {\bibfnamefont {P.}~\bibnamefont
  {Elliott}}, \bibinfo {author} {\bibfnamefont {K.}~\bibnamefont {Krieger}},
  \bibinfo {author} {\bibfnamefont {J.~K.}\ \bibnamefont {Dewhurst}}, \bibinfo
  {author} {\bibfnamefont {S.}~\bibnamefont {Sharma}}, \ and\ \bibinfo {author}
  {\bibfnamefont {E.~K.~U.}\ \bibnamefont {Gross}},\ }\href
  {http://stacks.iop.org/1367-2630/18/i=1/a=013014} {\bibfield  {journal}
  {\bibinfo  {journal} {New Journal of Physics}\ }\textbf {\bibinfo {volume}
  {18}},\ \bibinfo {pages} {013014} (\bibinfo {year} {2016})}\BibitemShut
  {NoStop}%
\bibitem [{\citenamefont {Raghunathan}\ and\ \citenamefont
  {Nest}(2011)}]{RN11}%
  \BibitemOpen
  \bibfield  {author} {\bibinfo {author} {\bibfnamefont {S.}~\bibnamefont
  {Raghunathan}}\ and\ \bibinfo {author} {\bibfnamefont {M.}~\bibnamefont
  {Nest}},\ }\href {\doibase 10.1021/ct200270t} {\bibfield  {journal} {\bibinfo
   {journal} {J. Chem. Theory and Comput.}\ }\textbf {\bibinfo {volume} {7}},\
  \bibinfo {pages} {2492} (\bibinfo {year} {2011})}\BibitemShut {NoStop}%
\bibitem [{\citenamefont {Raghunathan}\ and\ \citenamefont
  {Nest}(2012)}]{RN12c}%
  \BibitemOpen
  \bibfield  {author} {\bibinfo {author} {\bibfnamefont {S.}~\bibnamefont
  {Raghunathan}}\ and\ \bibinfo {author} {\bibfnamefont {M.}~\bibnamefont
  {Nest}},\ }\href@noop {} {\bibfield  {journal} {\bibinfo  {journal} {J. Chem.
  Theory and Comput.}\ }\textbf {\bibinfo {volume} {8}},\ \bibinfo {pages}
  {806} (\bibinfo {year} {2012})}\BibitemShut {NoStop}%
\bibitem [{\citenamefont {Habenicht}\ \emph {et~al.}(2014)\citenamefont
  {Habenicht}, \citenamefont {Tani}, \citenamefont {Provorse},\ and\
  \citenamefont {Isborn}}]{HTPI14}%
  \BibitemOpen
  \bibfield  {author} {\bibinfo {author} {\bibfnamefont {B.~F.}\ \bibnamefont
  {Habenicht}}, \bibinfo {author} {\bibfnamefont {N.~P.}\ \bibnamefont {Tani}},
  \bibinfo {author} {\bibfnamefont {M.~R.}\ \bibnamefont {Provorse}}, \ and\
  \bibinfo {author} {\bibfnamefont {C.~M.}\ \bibnamefont {Isborn}},\
  }\href@noop {} {\bibfield  {journal} {\bibinfo  {journal} {J. Chem. Phys.}\
  }\textbf {\bibinfo {volume} {141}},\ \bibinfo {pages} {184112} (\bibinfo
  {year} {2014})}\BibitemShut {NoStop}%
\bibitem [{\citenamefont {Ramsden}\ and\ \citenamefont {Godby}(2012)}]{RG12}%
  \BibitemOpen
  \bibfield  {author} {\bibinfo {author} {\bibfnamefont {J.}~\bibnamefont
  {Ramsden}}\ and\ \bibinfo {author} {\bibfnamefont {R.}~\bibnamefont
  {Godby}},\ }\href@noop {} {\bibfield  {journal} {\bibinfo  {journal} {Phys.
  Rev. Lett.}\ }\textbf {\bibinfo {volume} {109}},\ \bibinfo {pages} {036402}
  (\bibinfo {year} {2012})}\BibitemShut {NoStop}%
\bibitem [{\citenamefont {Hodgson}\ \emph {et~al.}(2013)\citenamefont
  {Hodgson}, \citenamefont {Ramsden}, \citenamefont {Chapman}, \citenamefont
  {Lillystone},\ and\ \citenamefont {Godby}}]{HRCLG13}%
  \BibitemOpen
  \bibfield  {author} {\bibinfo {author} {\bibfnamefont {M.~J.~P.}\
  \bibnamefont {Hodgson}}, \bibinfo {author} {\bibfnamefont {J.~D.}\
  \bibnamefont {Ramsden}}, \bibinfo {author} {\bibfnamefont {J.~B.~J.}\
  \bibnamefont {Chapman}}, \bibinfo {author} {\bibfnamefont {P.}~\bibnamefont
  {Lillystone}}, \ and\ \bibinfo {author} {\bibfnamefont {R.~W.}\ \bibnamefont
  {Godby}},\ }\href {\doibase 10.1103/PhysRevB.88.241102} {\bibfield  {journal}
  {\bibinfo  {journal} {Phys. Rev. B}\ }\textbf {\bibinfo {volume} {88}},\
  \bibinfo {pages} {241102} (\bibinfo {year} {2013})}\BibitemShut {NoStop}%
\bibitem [{\citenamefont {Hodgson}\ \emph {et~al.}(2014)\citenamefont
  {Hodgson}, \citenamefont {Ramsden}, \citenamefont {Durrant},\ and\
  \citenamefont {Godby}}]{HRDG14}%
  \BibitemOpen
  \bibfield  {author} {\bibinfo {author} {\bibfnamefont {M.~J.~P.}\
  \bibnamefont {Hodgson}}, \bibinfo {author} {\bibfnamefont {J.~D.}\
  \bibnamefont {Ramsden}}, \bibinfo {author} {\bibfnamefont {T.~R.}\
  \bibnamefont {Durrant}}, \ and\ \bibinfo {author} {\bibfnamefont {R.~W.}\
  \bibnamefont {Godby}},\ }\href {\doibase 10.1103/PhysRevB.90.241107}
  {\bibfield  {journal} {\bibinfo  {journal} {Phys. Rev. B}\ }\textbf {\bibinfo
  {volume} {90}},\ \bibinfo {pages} {241107} (\bibinfo {year}
  {2014})}\BibitemShut {NoStop}%
\bibitem [{\citenamefont {Elliott}\ \emph {et~al.}(2012)\citenamefont
  {Elliott}, \citenamefont {Fuks}, \citenamefont {Rubio},\ and\ \citenamefont
  {Maitra}}]{EFRM12}%
  \BibitemOpen
  \bibfield  {author} {\bibinfo {author} {\bibfnamefont {P.}~\bibnamefont
  {Elliott}}, \bibinfo {author} {\bibfnamefont {J.~I.}\ \bibnamefont {Fuks}},
  \bibinfo {author} {\bibfnamefont {A.}~\bibnamefont {Rubio}}, \ and\ \bibinfo
  {author} {\bibfnamefont {N.~T.}\ \bibnamefont {Maitra}},\ }\href {\doibase
  10.1103/PhysRevLett.109.266404} {\bibfield  {journal} {\bibinfo  {journal}
  {Phys. Rev. Lett.}\ }\textbf {\bibinfo {volume} {109}},\ \bibinfo {pages}
  {266404} (\bibinfo {year} {2012})}\BibitemShut {NoStop}%
\bibitem [{\citenamefont {Fuks}\ \emph {et~al.}(2013)\citenamefont {Fuks},
  \citenamefont {Elliott}, \citenamefont {Rubio},\ and\ \citenamefont
  {Maitra}}]{FERM13}%
  \BibitemOpen
  \bibfield  {author} {\bibinfo {author} {\bibfnamefont {J.~I.}\ \bibnamefont
  {Fuks}}, \bibinfo {author} {\bibfnamefont {P.}~\bibnamefont {Elliott}},
  \bibinfo {author} {\bibfnamefont {A.}~\bibnamefont {Rubio}}, \ and\ \bibinfo
  {author} {\bibfnamefont {N.~T.}\ \bibnamefont {Maitra}},\ }\href@noop {}
  {\bibfield  {journal} {\bibinfo  {journal} {J. Phys. Chem. Lett.}\ }\textbf
  {\bibinfo {volume} {4}},\ \bibinfo {pages} {735} (\bibinfo {year}
  {2013})}\BibitemShut {NoStop}%
\bibitem [{\citenamefont {Luo}\ \emph {et~al.}(2014)\citenamefont {Luo},
  \citenamefont {Fuks}, \citenamefont {Sandoval}, \citenamefont {Elliott},\
  and\ \citenamefont {Maitra}}]{LFSEM14}%
  \BibitemOpen
  \bibfield  {author} {\bibinfo {author} {\bibfnamefont {K.}~\bibnamefont
  {Luo}}, \bibinfo {author} {\bibfnamefont {J.~I.}\ \bibnamefont {Fuks}},
  \bibinfo {author} {\bibfnamefont {E.~D.}\ \bibnamefont {Sandoval}}, \bibinfo
  {author} {\bibfnamefont {P.}~\bibnamefont {Elliott}}, \ and\ \bibinfo
  {author} {\bibfnamefont {N.~T.}\ \bibnamefont {Maitra}},\ }\href@noop {}
  {\bibfield  {journal} {\bibinfo  {journal} {J. Chem. Phys}\ }\textbf
  {\bibinfo {volume} {140}},\ \bibinfo {pages} {18A515} (\bibinfo {year}
  {2014})}\BibitemShut {NoStop}%
\bibitem [{\citenamefont {Luo}\ \emph {et~al.}(2013)\citenamefont {Luo},
  \citenamefont {Elliott},\ and\ \citenamefont {Maitra}}]{LEM13}%
  \BibitemOpen
  \bibfield  {author} {\bibinfo {author} {\bibfnamefont {K.}~\bibnamefont
  {Luo}}, \bibinfo {author} {\bibfnamefont {P.}~\bibnamefont {Elliott}}, \ and\
  \bibinfo {author} {\bibfnamefont {N.~T.}\ \bibnamefont {Maitra}},\ }\href
  {\doibase 10.1103/PhysRevA.88.042508} {\bibfield  {journal} {\bibinfo
  {journal} {Phys. Rev. A}\ }\textbf {\bibinfo {volume} {88}},\ \bibinfo
  {pages} {042508} (\bibinfo {year} {2013})}\BibitemShut {NoStop}%
\bibitem [{\citenamefont {Buijse}\ \emph {et~al.}(1989)\citenamefont {Buijse},
  \citenamefont {Baerends},\ and\ \citenamefont {Snijders}}]{BBS89}%
  \BibitemOpen
  \bibfield  {author} {\bibinfo {author} {\bibfnamefont {M.~A.}\ \bibnamefont
  {Buijse}}, \bibinfo {author} {\bibfnamefont {E.~J.}\ \bibnamefont
  {Baerends}}, \ and\ \bibinfo {author} {\bibfnamefont {J.~G.}\ \bibnamefont
  {Snijders}},\ }\href {\doibase 10.1103/PhysRevA.40.4190} {\bibfield
  {journal} {\bibinfo  {journal} {Phys. Rev. A}\ }\textbf {\bibinfo {volume}
  {40}},\ \bibinfo {pages} {4190} (\bibinfo {year} {1989})}\BibitemShut
  {NoStop}%
\bibitem [{\citenamefont {Gritsenko}\ \emph {et~al.}(1994)\citenamefont
  {Gritsenko}, \citenamefont {van Leeuwen},\ and\ \citenamefont
  {Baerends}}]{GLB94}%
  \BibitemOpen
  \bibfield  {author} {\bibinfo {author} {\bibfnamefont {O.~V.}\ \bibnamefont
  {Gritsenko}}, \bibinfo {author} {\bibfnamefont {R.}~\bibnamefont {van
  Leeuwen}}, \ and\ \bibinfo {author} {\bibfnamefont {E.~J.}\ \bibnamefont
  {Baerends}},\ }\href {\doibase http://dx.doi.org/10.1063/1.468024} {\bibfield
   {journal} {\bibinfo  {journal} {The Journal of Chemical Physics}\ }\textbf
  {\bibinfo {volume} {101}},\ \bibinfo {pages} {8955} (\bibinfo {year}
  {1994})}\BibitemShut {NoStop}%
\bibitem [{\citenamefont {Gritsenko}\ \emph {et~al.}(1996)\citenamefont
  {Gritsenko}, \citenamefont {van Leeuwen},\ and\ \citenamefont
  {Baerends}}]{GLB96}%
  \BibitemOpen
  \bibfield  {author} {\bibinfo {author} {\bibfnamefont {O.~V.}\ \bibnamefont
  {Gritsenko}}, \bibinfo {author} {\bibfnamefont {R.}~\bibnamefont {van
  Leeuwen}}, \ and\ \bibinfo {author} {\bibfnamefont {E.~J.}\ \bibnamefont
  {Baerends}},\ }\href {\doibase http://dx.doi.org/10.1063/1.471602} {\bibfield
   {journal} {\bibinfo  {journal} {The Journal of Chemical Physics}\ }\textbf
  {\bibinfo {volume} {104}},\ \bibinfo {pages} {8535} (\bibinfo {year}
  {1996})}\BibitemShut {NoStop}%
\bibitem [{\citenamefont {Gritsenko}\ and\ \citenamefont
  {Baerends}(1996)}]{GB96}%
  \BibitemOpen
  \bibfield  {author} {\bibinfo {author} {\bibfnamefont {O.~V.}\ \bibnamefont
  {Gritsenko}}\ and\ \bibinfo {author} {\bibfnamefont {E.~J.}\ \bibnamefont
  {Baerends}},\ }\href {\doibase 10.1103/PhysRevA.54.1957} {\bibfield
  {journal} {\bibinfo  {journal} {Phys. Rev. A}\ }\textbf {\bibinfo {volume}
  {54}},\ \bibinfo {pages} {1957} (\bibinfo {year} {1996})}\BibitemShut
  {NoStop}%
\bibitem [{\citenamefont {Elliott}\ and\ \citenamefont {Maitra}(2012)}]{EM12}%
  \BibitemOpen
  \bibfield  {author} {\bibinfo {author} {\bibfnamefont {P.}~\bibnamefont
  {Elliott}}\ and\ \bibinfo {author} {\bibfnamefont {N.~T.}\ \bibnamefont
  {Maitra}},\ }\href@noop {} {\bibfield  {journal} {\bibinfo  {journal} {Phys.
  Rev. A}\ }\textbf {\bibinfo {volume} {85}},\ \bibinfo {pages} {052510}
  (\bibinfo {year} {2012})}\BibitemShut {NoStop}%
\bibitem [{\citenamefont {Nielsen}\ \emph {et~al.}(2013)\citenamefont
  {Nielsen}, \citenamefont {Ruggenthaler},\ and\ \citenamefont {van
  Leeuwen}}]{NRL13}%
  \BibitemOpen
  \bibfield  {author} {\bibinfo {author} {\bibfnamefont {S.}~\bibnamefont
  {Nielsen}}, \bibinfo {author} {\bibfnamefont {M.}~\bibnamefont
  {Ruggenthaler}}, \ and\ \bibinfo {author} {\bibfnamefont {R.}~\bibnamefont
  {van Leeuwen}},\ }\href@noop {} {\bibfield  {journal} {\bibinfo  {journal}
  {Europhys. Lett.}\ }\textbf {\bibinfo {volume} {101}},\ \bibinfo {pages}
  {33001} (\bibinfo {year} {2013})}\BibitemShut {NoStop}%
\bibitem [{\citenamefont {Nielsen}\ \emph {et~al.}(2014)\citenamefont
  {Nielsen}, \citenamefont {Ruggenthaler},\ and\ \citenamefont {van
  Leeuwen}}]{NRL14}%
  \BibitemOpen
  \bibfield  {author} {\bibinfo {author} {\bibfnamefont {S.}~\bibnamefont
  {Nielsen}}, \bibinfo {author} {\bibfnamefont {M.}~\bibnamefont
  {Ruggenthaler}}, \ and\ \bibinfo {author} {\bibfnamefont {R.}~\bibnamefont
  {van Leeuwen}},\ }\href@noop {} {\bibfield  {journal} {\bibinfo  {journal}
  {arXiv preprint arXiv:1412.3794}\ } (\bibinfo {year} {2014})}\BibitemShut
  {NoStop}%
\bibitem [{\citenamefont {Ruggenthaler}\ and\ \citenamefont {van
  Leeuwen}(2011)}]{RL11}%
  \BibitemOpen
  \bibfield  {author} {\bibinfo {author} {\bibfnamefont {M.}~\bibnamefont
  {Ruggenthaler}}\ and\ \bibinfo {author} {\bibfnamefont {R.}~\bibnamefont {van
  Leeuwen}},\ }\href@noop {} {\bibfield  {journal} {\bibinfo  {journal} {EPL
  (Europhysics Letters)}\ }\textbf {\bibinfo {volume} {95}},\ \bibinfo {pages}
  {13001} (\bibinfo {year} {2011})}\BibitemShut {NoStop}%
\bibitem [{\citenamefont {Ruggenthaler}\ \emph {et~al.}(2012)\citenamefont
  {Ruggenthaler}, \citenamefont {Giesbertz}, \citenamefont {Penz},\ and\
  \citenamefont {van Leeuwen}}]{RGPL12}%
  \BibitemOpen
  \bibfield  {author} {\bibinfo {author} {\bibfnamefont {M.}~\bibnamefont
  {Ruggenthaler}}, \bibinfo {author} {\bibfnamefont {K.}~\bibnamefont
  {Giesbertz}}, \bibinfo {author} {\bibfnamefont {M.}~\bibnamefont {Penz}}, \
  and\ \bibinfo {author} {\bibfnamefont {R.}~\bibnamefont {van Leeuwen}},\
  }\href@noop {} {\bibfield  {journal} {\bibinfo  {journal} {Physical Review
  A}\ }\textbf {\bibinfo {volume} {85}},\ \bibinfo {pages} {052504} (\bibinfo
  {year} {2012})}\BibitemShut {NoStop}%
\bibitem [{\citenamefont {Penz}\ and\ \citenamefont
  {Ruggenthaler}(2011)}]{PR11}%
  \BibitemOpen
  \bibfield  {author} {\bibinfo {author} {\bibfnamefont {M.}~\bibnamefont
  {Penz}}\ and\ \bibinfo {author} {\bibfnamefont {M.}~\bibnamefont
  {Ruggenthaler}},\ }\href@noop {} {\bibfield  {journal} {\bibinfo  {journal}
  {Journal of Physics A: Mathematical and Theoretical}\ }\textbf {\bibinfo
  {volume} {44}},\ \bibinfo {pages} {335208} (\bibinfo {year}
  {2011})}\BibitemShut {NoStop}%
\bibitem [{\citenamefont {Ruggenthaler}\ \emph {et~al.}(2015)\citenamefont
  {Ruggenthaler}, \citenamefont {Penz},\ and\ \citenamefont {van
  Leeuwen}}]{RPL15}%
  \BibitemOpen
  \bibfield  {author} {\bibinfo {author} {\bibfnamefont {M.}~\bibnamefont
  {Ruggenthaler}}, \bibinfo {author} {\bibfnamefont {M.}~\bibnamefont {Penz}},
  \ and\ \bibinfo {author} {\bibfnamefont {R.}~\bibnamefont {van Leeuwen}},\
  }\href@noop {} {\bibfield  {journal} {\bibinfo  {journal} {Journal of
  Physics: Condensed Matter}\ }\textbf {\bibinfo {volume} {27}},\ \bibinfo
  {pages} {203202} (\bibinfo {year} {2015})}\BibitemShut {NoStop}%
\bibitem [{\citenamefont {Thiele}\ \emph {et~al.}(2008)\citenamefont {Thiele},
  \citenamefont {Gross},\ and\ \citenamefont {K\"ummel}}]{TGK08}%
  \BibitemOpen
  \bibfield  {author} {\bibinfo {author} {\bibfnamefont {M.}~\bibnamefont
  {Thiele}}, \bibinfo {author} {\bibfnamefont {E.~K.~U.}\ \bibnamefont
  {Gross}}, \ and\ \bibinfo {author} {\bibfnamefont {S.}~\bibnamefont
  {K\"ummel}},\ }\href {\doibase 10.1103/PhysRevLett.100.153004} {\bibfield
  {journal} {\bibinfo  {journal} {Phys. Rev. Lett.}\ }\textbf {\bibinfo
  {volume} {100}},\ \bibinfo {pages} {153004} (\bibinfo {year}
  {2008})}\BibitemShut {NoStop}%
\bibitem [{\citenamefont {Ruggenthaler}\ and\ \citenamefont
  {Bauer}(2009)}]{RB09b}%
  \BibitemOpen
  \bibfield  {author} {\bibinfo {author} {\bibfnamefont {M.}~\bibnamefont
  {Ruggenthaler}}\ and\ \bibinfo {author} {\bibfnamefont {D.}~\bibnamefont
  {Bauer}},\ }\href@noop {} {\bibfield  {journal} {\bibinfo  {journal}
  {Physical Review A}\ }\textbf {\bibinfo {volume} {80}},\ \bibinfo {pages}
  {052502} (\bibinfo {year} {2009})}\BibitemShut {NoStop}%
\bibitem [{\citenamefont {Javanainen}\ \emph {et~al.}(1988)\citenamefont
  {Javanainen}, \citenamefont {Eberly},\ and\ \citenamefont {Su}}]{JES88}%
  \BibitemOpen
  \bibfield  {author} {\bibinfo {author} {\bibfnamefont {J.}~\bibnamefont
  {Javanainen}}, \bibinfo {author} {\bibfnamefont {J.~H.}\ \bibnamefont
  {Eberly}}, \ and\ \bibinfo {author} {\bibfnamefont {Q.}~\bibnamefont {Su}},\
  }\href {\doibase 10.1103/PhysRevA.38.3430} {\bibfield  {journal} {\bibinfo
  {journal} {Phys. Rev. A}\ }\textbf {\bibinfo {volume} {38}},\ \bibinfo
  {pages} {3430} (\bibinfo {year} {1988})}\BibitemShut {NoStop}%
\bibitem [{\citenamefont {van Leeuwen}\ and\ \citenamefont
  {Baerends}(1994)}]{VLB1994}%
  \BibitemOpen
  \bibfield  {author} {\bibinfo {author} {\bibfnamefont {R.}~\bibnamefont {van
  Leeuwen}}\ and\ \bibinfo {author} {\bibfnamefont {E.~J.}\ \bibnamefont
  {Baerends}},\ }\href {\doibase 10.1103/PhysRevA.49.2421} {\bibfield
  {journal} {\bibinfo  {journal} {Phys. Rev. A}\ }\textbf {\bibinfo {volume}
  {49}},\ \bibinfo {pages} {2421} (\bibinfo {year} {1994})}\BibitemShut
  {NoStop}%
\bibitem [{\citenamefont {Ruggenthaler}\ \emph {et~al.}(2013)\citenamefont
  {Ruggenthaler}, \citenamefont {Nielsen},\ and\ \citenamefont
  {Van~Leeuwen}}]{RNL13}%
  \BibitemOpen
  \bibfield  {author} {\bibinfo {author} {\bibfnamefont {M.}~\bibnamefont
  {Ruggenthaler}}, \bibinfo {author} {\bibfnamefont {S.~E.~B.}\ \bibnamefont
  {Nielsen}}, \ and\ \bibinfo {author} {\bibfnamefont {R.}~\bibnamefont
  {Van~Leeuwen}},\ }\href@noop {} {\bibfield  {journal} {\bibinfo  {journal}
  {Physical Review A}\ }\textbf {\bibinfo {volume} {88}},\ \bibinfo {pages}
  {022512} (\bibinfo {year} {2013})}\BibitemShut {NoStop}%
\bibitem [{\citenamefont {Schmitteckert}\ \emph {et~al.}(2013)\citenamefont
  {Schmitteckert}, \citenamefont {Dzierzawa},\ and\ \citenamefont
  {Schwab}}]{SDS13}%
  \BibitemOpen
  \bibfield  {author} {\bibinfo {author} {\bibfnamefont {P.}~\bibnamefont
  {Schmitteckert}}, \bibinfo {author} {\bibfnamefont {M.}~\bibnamefont
  {Dzierzawa}}, \ and\ \bibinfo {author} {\bibfnamefont {P.}~\bibnamefont
  {Schwab}},\ }\href {\doibase 10.1039/C3CP44639E} {\bibfield  {journal}
  {\bibinfo  {journal} {Phys. Chem. Chem. Phys.}\ }\textbf {\bibinfo {volume}
  {15}},\ \bibinfo {pages} {5477} (\bibinfo {year} {2013})}\BibitemShut
  {NoStop}%
\bibitem [{\citenamefont {Fuks}\ and\ \citenamefont {Maitra}(2014)}]{FM14}%
  \BibitemOpen
  \bibfield  {author} {\bibinfo {author} {\bibfnamefont {J.~I.}\ \bibnamefont
  {Fuks}}\ and\ \bibinfo {author} {\bibfnamefont {N.~T.}\ \bibnamefont
  {Maitra}},\ }\href {\doibase 10.1039/C4CP00118D} {\bibfield  {journal}
  {\bibinfo  {journal} {Phys. Chem. Chem. Phys.}\ }\textbf {\bibinfo {volume}
  {16}},\ \bibinfo {pages} {14504} (\bibinfo {year} {2014})}\BibitemShut
  {NoStop}%
\bibitem [{\citenamefont {Karlsson}\ \emph {et~al.}(2011)\citenamefont
  {Karlsson}, \citenamefont {Privitera},\ and\ \citenamefont
  {Verdozzi}}]{KPV11}%
  \BibitemOpen
  \bibfield  {author} {\bibinfo {author} {\bibfnamefont {D.}~\bibnamefont
  {Karlsson}}, \bibinfo {author} {\bibfnamefont {A.}~\bibnamefont {Privitera}},
  \ and\ \bibinfo {author} {\bibfnamefont {C.}~\bibnamefont {Verdozzi}},\
  }\href {\doibase 10.1103/PhysRevLett.106.116401} {\bibfield  {journal}
  {\bibinfo  {journal} {Phys. Rev. Lett.}\ }\textbf {\bibinfo {volume} {106}},\
  \bibinfo {pages} {116401} (\bibinfo {year} {2011})}\BibitemShut {NoStop}%
\bibitem [{\citenamefont {Verdozzi}\ \emph {et~al.}(2011)\citenamefont
  {Verdozzi}, \citenamefont {Karlsson}, \citenamefont {von Friesen},
  \citenamefont {Almbladh},\ and\ \citenamefont {von Barth}}]{VKFAB11}%
  \BibitemOpen
  \bibfield  {author} {\bibinfo {author} {\bibfnamefont {C.}~\bibnamefont
  {Verdozzi}}, \bibinfo {author} {\bibfnamefont {D.}~\bibnamefont {Karlsson}},
  \bibinfo {author} {\bibfnamefont {M.~P.}\ \bibnamefont {von Friesen}},
  \bibinfo {author} {\bibfnamefont {C.-O.}\ \bibnamefont {Almbladh}}, \ and\
  \bibinfo {author} {\bibfnamefont {U.}~\bibnamefont {von Barth}},\ }\href
  {\doibase http://dx.doi.org/10.1016/j.chemphys.2011.04.035} {\bibfield
  {journal} {\bibinfo  {journal} {Chemical Physics}\ }\textbf {\bibinfo
  {volume} {391}},\ \bibinfo {pages} {37 } (\bibinfo {year} {2011})},\ \bibinfo
  {note} {open problems and new solutions in time dependent density functional
  theory}\BibitemShut {NoStop}%
\bibitem [{\citenamefont {Tokatly}(2005)}]{T2005}%
  \BibitemOpen
  \bibfield  {author} {\bibinfo {author} {\bibfnamefont {I.~V.}\ \bibnamefont
  {Tokatly}},\ }\href {\doibase 10.1103/PhysRevB.71.165104} {\bibfield
  {journal} {\bibinfo  {journal} {Phys. Rev. B}\ }\textbf {\bibinfo {volume}
  {71}},\ \bibinfo {pages} {165104} (\bibinfo {year} {2005})}\BibitemShut
  {NoStop}%
\bibitem [{\citenamefont {G\"orling}(1999)}]{G99}%
  \BibitemOpen
  \bibfield  {author} {\bibinfo {author} {\bibfnamefont {A.}~\bibnamefont
  {G\"orling}},\ }\href {\doibase 10.1103/PhysRevLett.83.5459} {\bibfield
  {journal} {\bibinfo  {journal} {Phys. Rev. Lett.}\ }\textbf {\bibinfo
  {volume} {83}},\ \bibinfo {pages} {5459} (\bibinfo {year}
  {1999})}\BibitemShut {NoStop}%
\bibitem [{\citenamefont {Baerends}\ and\ \citenamefont
  {Gritsenko}(2005)}]{BG05}%
  \BibitemOpen
  \bibfield  {author} {\bibinfo {author} {\bibfnamefont {E.~J.}\ \bibnamefont
  {Baerends}}\ and\ \bibinfo {author} {\bibfnamefont {O.~V.}\ \bibnamefont
  {Gritsenko}},\ }\href {\doibase http://dx.doi.org/10.1063/1.1904566}
  {\bibfield  {journal} {\bibinfo  {journal} {The Journal of Chemical Physics}\
  }\textbf {\bibinfo {volume} {123}},\ \bibinfo {eid} {062202} (\bibinfo {year}
  {2005}),\ http://dx.doi.org/10.1063/1.1904566}\BibitemShut {NoStop}%
\end{thebibliography}%


\end{document}